\documentclass[aip,jcp,superscriptaddress,reprint,showpacs,amsfonts,amsmath]{revtex4-1}
\usepackage[final]{graphicx}
\usepackage{comment}
\usepackage[normalem]{ulem}
\usepackage{xfrac}
\usepackage{upgreek}
\usepackage{units}

\let\bs\boldsymbol
\DeclareMathOperator\tr{tr}
\newcommand{\tlname}[1]{\ensuremath{\text{\textit{#1}}}}

\usepackage[usenames,dvipsnames]{xcolor}
\usepackage[normalem]{ulem}

\newcommand{\twall}{t_\text{wall}}

\renewcommand{\vec}[1]{\boldsymbol{#1}}

\begin{document}
\title{Lattice Boltzmann simulations of a viscoelastic shear-thinning fluid}
\date\today
\newcommand\dlr{\affiliation{Institut f\"ur Materialphysik im Weltraum,
  Deutsches Zentrum f\"ur Luft- und Raumfahrt (DLR), 51170 K\"oln,
  Germany}}
\newcommand\hhu{\affiliation{Department of Physics,
  Heinrich-Heine Universit\"at D\"usseldorf,
  Universit\"atsstr.~1, 40225 D\"usseldorf, Germany}}
\author{S.~Papenkort}
\email{simon.papenkort@dlr.de}
\dlr
\author{Th.~Voigtmann}\dlr\hhu

\begin{abstract}
  We present a hybrid lattice Boltzmann algorithm for the simulation of
  flow glass-forming fluids, characterized by slow structural relaxation,
  at the level of the Navier-Stokes equation.
  The fluid is described in terms of a nonlinear integral constitutive equation,
  relating the stress tensor locally to the history of flow.
  As an application, we present results for an integral nonlinear Maxwell
  model that combines the effects of (linear) viscoelasticity and
  (nonlinear) shear thinning. We discuss the transient dynamics of velocities,
  shear stresses, and normal stress differences in planar pressure-driven
  channel flow, after switching on (startup) and off (cessation) of the driving
  pressure. This transient dynamics depends nontrivially on the channel
  width due to an interplay between hydrodynamic momentum diffusion and
  slow structural relaxation.
\end{abstract}

\maketitle

\section{Introduction}

The flow of glass forming fluids is characterized by an interplay of
slow, collective structural relaxation and flow-induced relaxation.
In many applications in particular in colloidal suspensions,
the structural relaxation rate $1/\tau$
is much larger than the flow rate $\dot\gamma$;
hence the relevant P\'eclet number $\tlname{Pe}=\dot\gamma\tau\gg1$. This leads
to pronounced nonlinear-response effects such as shear-thinning: the effective
viscosity of the fluid decreases strongly with increasing $\dot\gamma$.
This regime opens close to the glass transition, where $\tau\to\infty$,
and hence $\tlname{Pe}\gg1$ even if the shear rate is still
slow compared to the individual-particle short-time relaxaton rate $1/\tau_0$,
such that the bare P\'eclet number
$\tlname{Pe}_0=\dot\gamma\tau_0\ll1$. This is the regime of
nonlinear glassy rheology \cite{Voigtmann14cocis}.

In transient dynamics, viscoelastic and shear-thinning effects combine:
on time scales $t\gg\tau_0$ but $t\ll1/\dot\gamma$, the system
effectiely behaves as a transiently frozen amorphous structure
characterized by some elastic modulus $G_\infty$.
This was first recognized by Maxwell \cite{Maxwell1867} for the linear-response
regime ($\tlname{Pe}\ll1$). The simplest fluid model describing this
phenomenon, called Maxwell model in attribution to him, can be written as
\begin{equation}\label{maxwell}
  \sigma_{xy}(t)=\int_0^t\dot\gamma(t')G_\infty e^{-(t-t')/\tau}\,dt'\,,
\end{equation}
where simple-shear flow $\dot\gamma=\partial_yv_x$
is assumed to start at $t=0$ from a stress-free equilibrated state,
and $\sigma_{xy}(t)$ is the
shear-stress component of the Cauchy stress tensor $\boldsymbol\sigma(t)$.
For $t\ll\tau$, the model gives the stress-strain relation of an elastic
Hookean solid, $\sigma(t)\sim G_\infty\gamma_{t0}$ where
$\gamma_{tt'}=\int_{t'}^t\dot\gamma(s)\,ds$ is the accumulated shear strain.
For $t\gg\tau$ and constant shear rate, $\sigma(t)\sim\eta\dot\gamma$
recovers viscous Newtonian flow with a shear viscosity $\eta=G_\infty\tau$
given by the so-called Maxwell relation.

Equations such as Eq.~\eqref{maxwell} are constitutive equations for
continuum mechanics: the
macroscopic flow field $\vec v$ and its gradients $\boldsymbol\kappa
=(\vec\nabla\vec v)^T$ are determined by the Navier-Stokes equation
\cite{Salencon.2001},
\begin{equation}\label{ns}
  \partial_t\varrho\vec v+\vec\nabla\cdot(\varrho\vec v\vec v)
  =\varrho\vec f-\vec\nabla p+\vec\nabla\cdot\boldsymbol\sigma\,,
\end{equation}
where $\vec f$ is a given external force density, and $p$ is the thermodynamic
pressure. The mass-density field $\varrho$ obeys a continuity equation,
$\partial_t\varrho+\vec\nabla\cdot\varrho\vec v=0$. We assume in the following
incompressible flow, $\vec\nabla\cdot\vec v=0$.
The stress tensor $\boldsymbol\sigma$ expresses microscopic friction effects
that need to be described by a constitutive equation
that expresses $\boldsymbol\sigma$ in terms of
the flow fields again.

Recent theoretical work based on a formalism called integration through
transients (ITT), developed by Fuchs and Cates in the context of driven
colloidal fluids \cite{fuchs02,fuchs09}, gives a framework to derive
constitutive equations
for nonlinear glassy rheology from microscopic theory. Assuming flow
to remain homogeneous at least on mesoscopic scales, one gets, schematically,
\begin{equation}\label{gitt}
  \boldsymbol\sigma_\text{struc}(t)=\int_0^t[-\partial_{t'}\boldsymbol B(t,t')]
  G(t,t',[\boldsymbol\kappa])\,dt'\,,
\end{equation}
where the subscript ``struc'' recalls that this is only the structural-relaxation
contribution to the stresses (akin to the purely polymeric stress contribution
in polymer rheology).
$G(t,t',[\boldsymbol\kappa])$ is a nonlinear dynamical shear modulus that
depends on the
whole history of deformation gradients in a suitable way that honors
the invariance of stresses under rigid-body motions (called material-frame
indifference in continuum mechanics).
Replacing it with a flow-independent exponential recovers the Maxwell model.
$\boldsymbol B(t,t')=\boldsymbol E(t,t')\cdot\boldsymbol E^T(t,t')$ is
the Finger tensor, related to the deformation tensor
\begin{equation}
  \boldsymbol E(t,t')=\exp_+\left[\int_{t'}^t\boldsymbol\kappa(s)\,ds\right]\,.
\end{equation}
Here $\exp_+$ denotes a time-ordered exponential where all products of
$\boldsymbol\kappa$ are sorted such that earlier times appear to the right.
The appearance of $-\partial_{t'}\boldsymbol B(t,t')$ in Eq.~\eqref{gitt}
generalizes the scalar shear rate in Eq.~\eqref{maxwell} to arbitrary
incompressible flow geometries.

A natural feature of constitutive equations such as 
Eq.~\eqref{gitt} is their integral nature. Other than Eq.~\eqref{maxwell},
it cannot in general be reduced to a differential equation involving
only time-local differential operators. This is in particular relevant
close to the kinetic glass transition,
where the dominant feature in the dynamics
is the slow structural relaxation that provides long-lasting memory effects.
Within the mode-coupling theory of the glass transition (MCT), the
dynamical shear modulus is
determined from the solutions of a set of integro-differential equations
for density correlation functions that contain the effects of flow advection
\cite{fuchs09}.
As a key feature, the history integral does not have a predefined
natural cutoff, but rather it extends arbitrarily far back in time,
in a way that depends on the state point and the flow history. This
in essence reflects a lack of separation of time scales into short
microscopic ones and potentially slow hydrodynamic ones.

To study the flow of glass-forming fluids, it is hence desirable to develop
continuum-mechanics solvers that can combine the Navier-Stokes equation with
integral-type constitutive equations such as Eq.~\eqref{gitt}.
In this paper, we describe such a scheme, based on the lattice Boltzmann (LB)
algorithm \cite{Succi,Duenweg.2009} to solve the Navier-Stokes equation
in the low-Mach number limit.

The LB algorithm is a kinetic scheme based on lattice-node densities that
evolve according to collision-and-streaming rules taylored to reproduce
in the continuum limit of vanishing lattice spacing and time-step length
the Navier-Stokes equation for a Newtonian fluid. We base our work on
our recent LB algorithm incorporating non-Newtonian stresses
through a modified collision rule \cite{LBpaper}. This algorithm is here
extended to a full ``hybrid-LB'' scheme combining LB steps for the Navier-Stokes
equation with an integral-equation solver keping track of the full flow
history in Euler coordinates.
Previous approaches extending LB by additional lattice populations or
forcing terms have
focused on linear viscoelasticity \cite{giraud97,giraud98} and nonlinear
models that can be expressed in terms of differential constitutive equations
\cite{Denniston.2001,lallemand03,Sulaiman.2006,Malaspinas.2010,Malaspinas}; for a review see also
Ref.~\onlinecite{phillips11}. Also, modified collision rules can be used
to simulate the nonlinear rheology of emulsions, exploring the flexibility
of the LB algorithm as a kinetic scheme \cite{Benzi.2010,sbragaglia12,Benzi.2013}.
Earlier hybrid-LB schemes have coupled the LB algorithm with finite-difference
solvers for differential constitutive equations, usually entering non-Newtonian stresses in terms of a body force \cite{Marenduzzo.2007b,Henrich.2010,Frantziskonis.2011,su13}.
Our hybrid-LB scheme is new in its focus on integral constitutive equations
with large relaxation times.
For the treatment of integral constitutive equations combined with
finite-element and finite-volume
algorithms, see Refs.~\onlinecite{keunings03,tome08};
for a recent LB-based finite-volume formulation, see also Ref.~\onlinecite{bergamasco13}.

As a specific application, we implement a nonlinear generalization of
the Maxwell model that combines the effects of shear-thinning and
viscoelasticity. We use this model to study generic transient effects
of the startup and cessation of pressure-driven flow in a planer (2D) channel.

The paper is organized as follows: we introduce in Sec.~\ref{sec:maxwell}
the integral nonlinear Maxwell model.
In Sec.~\ref{sec:lb}, we briefly describe the LB algorithm based on
Ref.~\onlinecite{LBpaper} and our integral solver.
Section~\ref{sec:results} presents results for pressure-driven channel flow,
followed by a concluding Sec.~\ref{sec:concl}.

\section{Nonlinear Maxwell Model}\label{sec:maxwell}

The key interplay of mechanisms expressed by the ITT-MCT
\cite{fuchs02,fuchs09,brader07,brader08,brader09} is that of
slow structural relaxation of rate $\tau^{-1}$,
and of flow-induced relaxation of rate $|\dot\gamma|/\gamma_c$ (where
we introduce $\gamma_c$ as a model parameter controlling the effectiveness
of strain in breaking nearest-neighbor cages).
If $\tlname{Pe}\gg1\gg\tlname{Pe}_0$, the integral in Eq.~\eqref{gitt}
is cut off times $t'<t-\gamma_c/|\dot\gamma|$, leading to a decrease
of the effective viscosity, i.e., shear thinning.

In the Maxwell model, this can be incorporated by letting (in steady state)
$\tau^{-1}\mapsto\tau^{-1}+|\dot\gamma|/\gamma_c$. For general time-dependent
incompressible flow,
a plausible choice is to set, in Eq.~\eqref{gitt},
\begin{equation}\label{nlmax}
  G(t,t',[\boldsymbol\kappa])=G_\infty e^{-(t-t')/\tau}\,e^{-(t-t')
  |\dot\gamma(t')|/\gamma_c}\,dt'\,.
\end{equation}
For the flow rate, we set $\dot\gamma^2=\tlname{II}_D=(1/2)\tr\bs D^2$
(for incompressible flow), where $\bs D=\bs\kappa+\bs\kappa^T$ is the
symmetric velocity-gradient tensor.
The quantity $\gamma_c$ is set to $1/10$ in our numerical calculations;
this follows previous ITT-MCT investigations of colloidal glass formers
\cite{Amann.2013} where it was related to the typical fraction of a particle
diameter each particle can be sheared before cages break (agreeing with
an empirical criterion for the melting of solids given by Lindemann).
Our integral nonlinear Maxwell model (nlM) reproduces qualitative features
found in experiment and ITT-MCT for time-dependent
nonlinear glassy rheology, e.g., for
large-amplitude oscillatory shear or creep under imposed stress
\cite{siebenbuerger12,ThVcreep}. It also describes the discontinuous
emergence of a finite yield stress at the glass transition, $\tau\to\infty$.
It should however be kept in mind that Eq.~\eqref{nlmax} represents a
gross oversimplification of the ITT-MCT dynamics.
The identification of $\tlname{II}_D^{1/2}$ as
the local shear rate is ad-hoc. It is the simplest material-frame
indifferent choice that is linear and respects the symmetry of flow
reversal (under which the relaxation rate must remain positive).
A notable feature of our model that is faithful to the microscopic
ITT-MCT constitutive equation, is that the dynamical shear modulus defined by
Eq.~\eqref{nlmax}, breaks time-translational invariance for non-stationary flow.
For this reason it cannot be reduced to a differential constitutive equation.

Equation~\eqref{nlmax} together with Eq.~\eqref{gitt} reduce to the
well-known upper-convected Maxwell model (UCM) \cite{Larson} in the case where
$\tlname{Pe}\ll1$ (such that the second exponential in Eq.~\eqref{nlmax}
can be approximated by unity).
Although the UCM can be written in differential form, it will serve as
a useful test of our algorithm since analytical solutions are available
for the transient dynamics (see Appendix~\ref{sec:appendix}).
A different generalization of the UCM to nonlinear rheology in terms of
a differential equation is the
White-Metzner model \cite{whiteMetzner63}; it corresponds to replacing
the shear-rate-dependent expression in the second exponential of
Eq.~eqref{nlmax} by the accumulated strain. Which is closer
to true ITT-MCT may actually depend on the type of flow considered.
ITT-MCT keeps a much more complicated strain dependence that is not
easily cast into a simple form for the shear modulus.

The Maxwell model only addresses structural relaxation on times
large compared to $\tau_0$. On this short-time scale, the dynamical
shear modulus decays from its instantaneous value to the Maxwell plateau
modulus $G_\infty$. We are not concerned with this regime here, and
simply assume this process to provide a Newtonian background viscosity
$\eta_\infty=G_\infty\tau_0$ that is shear-rate independent. For colloidal
suspensions, this may be thought of as a crude model of the solvent
viscosity, ignoring the flow-induced hydrodynamic interaction effects.
We hence set
\begin{equation}
\label{nlM}
\bs\sigma(t) = \bs\sigma_\text{struc}(t) + \eta_\infty \bs D(t),
\end{equation}

\section{Lattice Boltzmann method}\label{sec:lb}

The lattice Boltzmann method is a fast and versatile tool to solve the
Navier-Stokes equations. For a review, we refer to
Ref.~\onlinecite{Duenweg.2009}.
Considering a regular rectangular spatial grid with lattice spacing $\delta x$,
the LB scheme evolves a set of lattice-density distributions $n_i$
are associated to a finite number of velocities $\vec c_i$ that represent
streaming from a node to, usually, the nearest and next-to-nearest
neighbors. The lattice distributions are evolved over a time step $\delta t$
by a collision-and-streaming rule
\begin{multline}
n_i(\vec r+\vec c_i\delta t, t+\delta t)=n_i^*(\vec r, t)\\ = n_i(\vec r, t)+\Delta_i[n(\vec r, t)] + F_i\,.
\end{multline}
The collision operator $\Delta_i$ implements relaxation towards a set of
equilibrium distributions $n_i^\text{eq}$ that are chosen such that for a
specific lattice, the desired continuum limit emerges as
$\delta x,\delta t\to0$. The term $F_i$ is used to model external forces or,
in our case, a non-Newtonian part of the stress tensor.

We consider a two-dimensional grid for simplicity; the extension
to 3D is straightforward. The velocity set is chosen according to the standard
D2Q9 model incorporating
nine lattice velocities: $\vec c_0=(0,0)$,
$\vec c_{1\ldots4}=(\pm 1, 0)c, (0, \pm 1)c$,
and $\vec c_{5\ldots8}\delta t=(\pm 1, \pm 1)c$, in units of the lattice
velocity $c=\delta x/\delta t$.
For the collision operator, a
single-relaxation time BGK model is employed,
\begin{align}
\Delta_i = -\frac 1 {\tau_\text{LB}} \left(n_i-n_i^\text{eq}\right)
\end{align}
where the equilibrium distributions are given by
\begin{align}
n_i^\text{eq}&(\rho, \vec u) = \nonumber\\ &a^{c_i}\rho\left(1+\frac {\vec u \cdot \vec c_i}{c_\text s^2}+\frac{\vec (c_{i\alpha} c_{i\beta} -c_\text s^2 \delta_{\alpha\beta})u_{i\alpha}u_{i\beta}}{2c_\text s^4}\right)\,.
\end{align}
The lattice weights $a^0=4/9$,
$a^1=1/9$, $a^{\sqrt 2}=1/36$, and the speed of sound $c_\text s=c/\sqrt3$
are chosen to reproduce the flow of a Newtonian fluid if the forcing term
$F_i$ is set to zero. In this case, the Newtonian shear viscosity is
given by $\eta_\text N=(\delta t)\rho c_\text s^2(\tau_\text{LB}-1/2)$.
Note that the equilibrium distribution depends only on
the fluid density $\rho$ and velocity $\vec u$, but not on the flow rate.

To model non-Newtonian stresses $\boldsymbol\sigma^\text{nN}$, we set
\cite{LBpaper}
\begin{multline}
\label{bodyforce}
  F_i=a^{c_i}\bigg\{\frac{-1}{2c_s^4\tau_\text{LB}}\bar\sigma_{\alpha\beta}
  ^\text{nN}
  (c_{i\alpha}c_{i\beta}-c_s^2\delta_{\alpha\beta})\\
+\Big(1-\frac 1 {2\tau_\text{LB}}\Big)\Big[(\delta t)(\partial_t\delta\rho)+\frac{f_\alpha^\text{ex}
  c_{i\alpha}}{c_s^2}+\frac{c_{i\alpha}c_{i\beta}-c_s^2\delta_{\alpha\beta}}{2c_s^4}\times\\
\Big(-(\delta t)(\partial_t\delta\rho)u_\alpha
    u_\beta+(u_\beta f_\alpha^\text{ex} + u_\alpha f_\beta^\text{ex}
  )\Big)\Big]\bigg\}\,.
\end{multline}
As the non-Newtonian stresses are not in general traceless, we split
$\bs\sigma^\text{nN}$ into its
traceless part denoted by an overbar, and a non-Newtonian pressure
contribution $\delta p^{\text{nN}}=-1/2\sigma_{\gamma\gamma}^\text{nN}$,
where $\sigma_{\gamma\gamma}^\text{nN}$ denotes the trace of the non-Newtonian
stress tensor. In incompressible flow, the
equation of the state of the D2Q9 model, $p_0=\rho c_\text s^2$, relates
this extra pressure term to a small variation in the density which is
implemented via
\begin{align}
\partial_t\delta\rho(t)=-\frac 1 {2 c_\text s ^2 \delta t}\left(\sigma_{\gamma\gamma}^\text{nN}(t)-\sigma_{\gamma\gamma}^\text{nN}(t-\delta t)\right).
\end{align}
The hydrodynamic density and momentum fields are recovered from the
relations \cite{LBpaper}
\begin{subequations}\label{hydrofields}
\begin{gather}
\rho(\vec r, t)=\sum_i n_i^\text{eq} = \sum_i n_i + \frac {\delta t}2 \partial_t\rho\,,\\
\rho \vec u(\vec r, t) = \sum_i \vec c_i n_i^\text{eq} = \sum_i \vec c_i n_i + \frac {\delta t} 2  \vec f\,.
\end{gather}
\end{subequations}
For the justification of this LB scheme, we refer to Ref.~\onlinecite{LBpaper},
where a standard Chapman-Enskog expansion was used to demonstrate
that the continuum limit of our scheme is indeed the Navier-Stokes equation
supplemented with a non-Newtonian stress contribution.

For the application to planar channel flow considered below, we assume
the flow to remain translational invariant for computational efficency.
Generalized periodic boundary conditions \cite{kim07} are used to fix a
constant pressure step between the inlet and outlet of the periodic lattice.
Under these conditions, only the terms quadratic in the velocities need to
be kept in Eq.~\eqref{bodyforce}. The definitions of the hydrodynamic fields,
Eq.~\eqref{hydrofields}, then reduce to the standard ones discussed in the
LB literature \cite{Duenweg.2009}.

To implement constitutive equations of the form of Eq.~\eqref{gitt},
we keep track of the Finger tensor $\bs B$ during the simulation, introducing
a time-history grid for the integration of the constitutive equation at each
LB lattice node.
Since in the regime of interest for nonlinear glassy rheology, the
structural relaxation time $\tau$ can be orders of magnitude larger than
$\tau_0$, the time integral in Eq.~\eqref{gitt} extends backwards over
a potentially large time span. To deal with this, we employ a
quasi-logarithmic memory layout consisting of $B$ blocks (labeled $b=1,\ldots B$)
of equidistant lattices with $C$ grid points of
fixed time step $\delta t_b$. The time steps are doubled from block to
block, as time extends backwards from $t$ to $t'<t$ in the integration:
$\delta t_{b-1}=2\delta t_b$, identifying the smallest step size as
that of the LB solver, $\delta t_B=\delta t$.
This quasi-logarithmic grid assumes that the function $G(t,t',[\bs\kappa])$
entering Eq.~\eqref{gitt} varies slowly for large $t-t'$; this is indeed
a feature of both our nlM model and the full ITT-MCT if the time-dependence
of the flow rate itself is not fast.

Instead of computing the time derivative of the Finger tensor
directly, we save the velocity gradient tensor $\bs \kappa$ and
the discrete contributions $\exp(\bs\kappa \delta t)$ to the deformation
tensor
\begin{multline}\label{eeval}
\bs E(t, t^\prime) = 
  e^{\bs \kappa(t-\delta t) \delta t}\ldots
  \underbrace{e^{\bs\kappa(t-(C-1)\delta t)\delta t}
  e^{\bs \kappa(t-C\delta t)\delta t }}
  _{e ^{\bs \kappa\delta t_{B-1}}} \\ 
  \ldots e ^{\bs \kappa (t^\prime)\delta t_{B-X}}\, . 
\end{multline}
As the LB scheme steps forward in time and a time-integration block is
filled, the two oldest entries are multiplied and moved to the
underlying block with a time step twice as big. Within lattice accuracy,
this procedure keeps the exact value of the deformation tensor.
For the velocity-gradient tensor $\bs\kappa$ a further approximation
is needed: we keep the averaged tensor over the two oldest
entries when transferring them to the next block backwards. The time
derivative of the Finger tensor is then evaluated according to
\begin{multline}
-\partial_t \bs B(t, t')= \bs E(t, t')\bs\kappa(t')\bs E^T(t, t')\\
  + (\bs E(t, t')\bs\kappa(t')\bs E^T(t, t'))^T\,.
\end{multline}

At each LB lattice node, the integration of Eq.~\eqref{gitt} can then be
performed by a suitable integration scheme. For the slowly varying functions
we expect on physical grounds, a simple trapezoidal rule is sufficient.
Our hybrid-LB scheme is particularly adapted to these situations where
the constitutive equation is given in terms of Euler coordinates (as
exemplified by the appearance of the Finger tensor, instead of the
Cauchy-Green tensor). Thus, we do not need to keep track of the flow-advected
movement of Lagrangian material points; at the expense of needing to
evaluate the time-ordered exponential Eq.~\eqref{eeval} based on the
generator $\bs\kappa$ of the nonlinear deformation.

In the following we will test the scheme for a simple 2D setup that is
translation-invariant in the flow direction. Assuming the flow to
remain laminar, we considerably increase the efficiency of the
computationally demanding algorithm (approximately $10^5$ lattice-node
updates per second using 1 core of an Intel Core i5-3470S CPU) by
evaluating the non-Newtonian stress only in the central column of the
lattice and relaying this extra stress along the symmetry axis. We
have checked in separate simulations that this does not affect the
results. Unless stated otherwise, we use $\tau_\text{LB}=0.9$ and a
grid of $100\times 20$ lattice nodes (depending on the Maxwell
relaxation time $\tau$). For the integration of the flow history,
we have chosen a block size
of $C=128$.
The velocity gradient tensor
$\bs\kappa$ is evaluated using a second-order finite difference
scheme.
The algorithm is implemented in the open-source lattice
Boltzmann code Palabos \cite{palabos}.

For the simulations, we choose parameters as follows: a pressure drop
along a channel of length $L$ is considered that is comparable to the
Maxwell elastic modulus, $\Delta p/L=G_\infty$; this is typical for
soft-matter fluids, where $G_\infty=O(1\,\text{Pa})$. Our model is that of a
yield-stress fluid, and this pressure difference is sufficient to
flow-melt the glassy state modeled by $\tau=\infty$.
Times are
measured in units of the microscopic relaxation time $\tau_0$ (on the
order of $1\,\text{ms}$ for typical colloidal fluids); the relevant
parameter expressing the viscoelasticity of the system is then the
relative slowdown of structural relaxation,
$\theta=\tau/\tau_0$. Unless stated otherwise, calculations are
performed for $\theta=10$. The dynamics does not change qualitatively
for higher values of $\theta$.


\section{Results}\label{sec:results}

\subsection{Steady State}

\begin{figure}
\includegraphics[width=.9\linewidth]{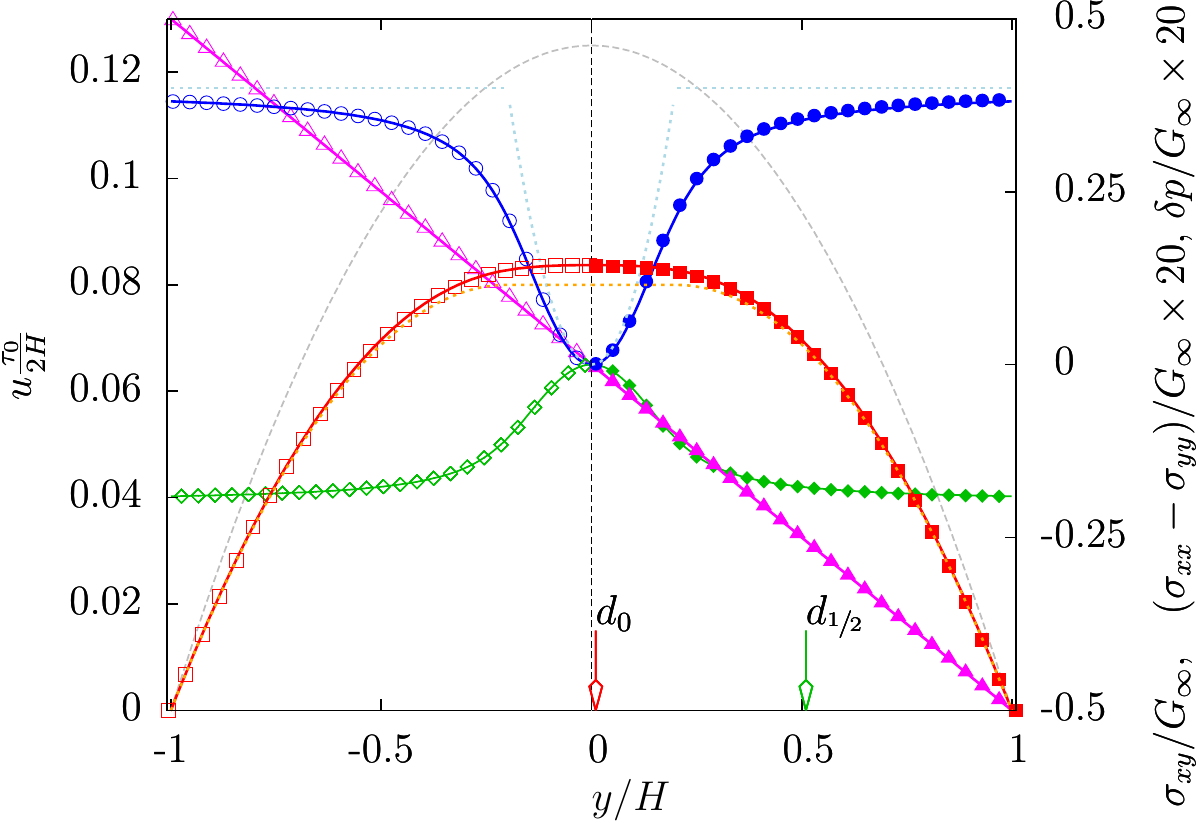}
\caption{\label{steadystate}Profiles of the steady-state velocity (boxes; left
  $y$-axis), shear stress (triangles), normal-stress difference
  (circles, multiplied by $20$), and pressure (diamonds, multiplied by $20$;
  right $y$-axis) for flow in a 2D channel of width $2H$
  (driven by a pressure gradient $\Delta
  p/(2H)=G_\infty$), for the nonlinear Maxwell model, Eq.~\eqref{nlmax},
  with $\tau/\tau_0=10$.
  Open symbols (left half) are for an instantaneous model,
  Ref.~\onlinecite{LBpaper}, closed symbols (right half) for the present model,
  obtained from our LB algorithm; lines are analytical solutions.
  The Newtonian (dashed line) and glassy (dotted line)
  profiles are shown for comparison.
  Arrows indicate the positions used in later plots.
}
\end{figure}

We briefly discuss the stationary flow profiles. For translational-invariant
flow, the model defined by Eqs.~\eqref{nlmax} and \eqref{ns} can be
solved analytically to give
\begin{subequations}
\begin{equation}
\vec\sigma_\text{ss}=\eta_\infty\left(\vec\kappa+\vec\kappa^T\right)+\sum_{n\ge 1} G_\infty \tau_\text{M}^n\vec d^{(n)},
\end{equation}
with
\begin{equation}
\vec d^{(n)}:=\sum_{m=0}^n {n \choose m} \vec \kappa^m \cdot {\vec \kappa^T}^{n-m}\,.
\end{equation}
\label{inlM}
\end{subequations}
Rereading this as stress--strain-rate relation at arbitrary time $t$ defines
the instantaneous nonlinear Maxwell model used in Ref.~\onlinecite{LBpaper}.
Such instantaneous constitutive equations (that relate the stresses
$\boldsymbol\sigma(t)$ to the strain rates $\boldsymbol\kappa(t)$ at the
same time) are numerically much less demanding to implement. However,
they only account for steady-state shear thinning and not
for viscoelasticity. To check that the integration scheme used here for
the flow-history integral converges to the correct steady state, we compare
in Fig.~\ref{steadystate} the velocity, shear-stress, and
normal-stress profiles of the two models with the analytical solution
for pressure-driven channel flow given in Ref.~\onlinecite{LBpaper}.
Both LB results match the analytical solution perfectly for the value
of $\tau=10\tau_0$. This value is chosen as a moderately
viscoelastic case (keeping the numerical effort in solving the history
integrals moderate), that already represents many features of the glassy
solution ($\tau\to\infty$, shown as dotted lines in Fig.~\ref{steadystate}).

Figure~\ref{steadystate} demonstrates typical effect for the channel
flow of viscoelastic shear-thinning fluids: for pressure gradients per
unit length comparable to the shear modulus, it is advantageous for
the fluid to form high-shear regions near the walls and a co-moving
low-shear ``plug'' in the center. This is explained by the finite yield
stress that arises at the glass transition. In the center of the channel,
the shear stress remains below the yield stress, so that no homogeneous
flow is possible there. For our model, the yield stress has a shear-stress
component of
$\sigma_{\text{y},xy}=G_\infty\gamma_c$.
The tensorial structure of the model includes normal stresses (circles
and diamonds in the figure). They are quadratic in the shear-rate due
to material-frame indifference of the stress tensor. In the flowing region
outside the plug, the model predicts a positive normal stress difference,
$\sigma_{xx}-\sigma_{yy}>0$. This causes a force towards the channel center,
which in the incompressible fluid is balanced by an increase in the
local density. In colloidal suspensions, this can be a driving mechanism
for particle migration to the center \cite{Nott.1994}.

\subsection{Transient dynamics}

\begin{figure}
\includegraphics[width=\linewidth]{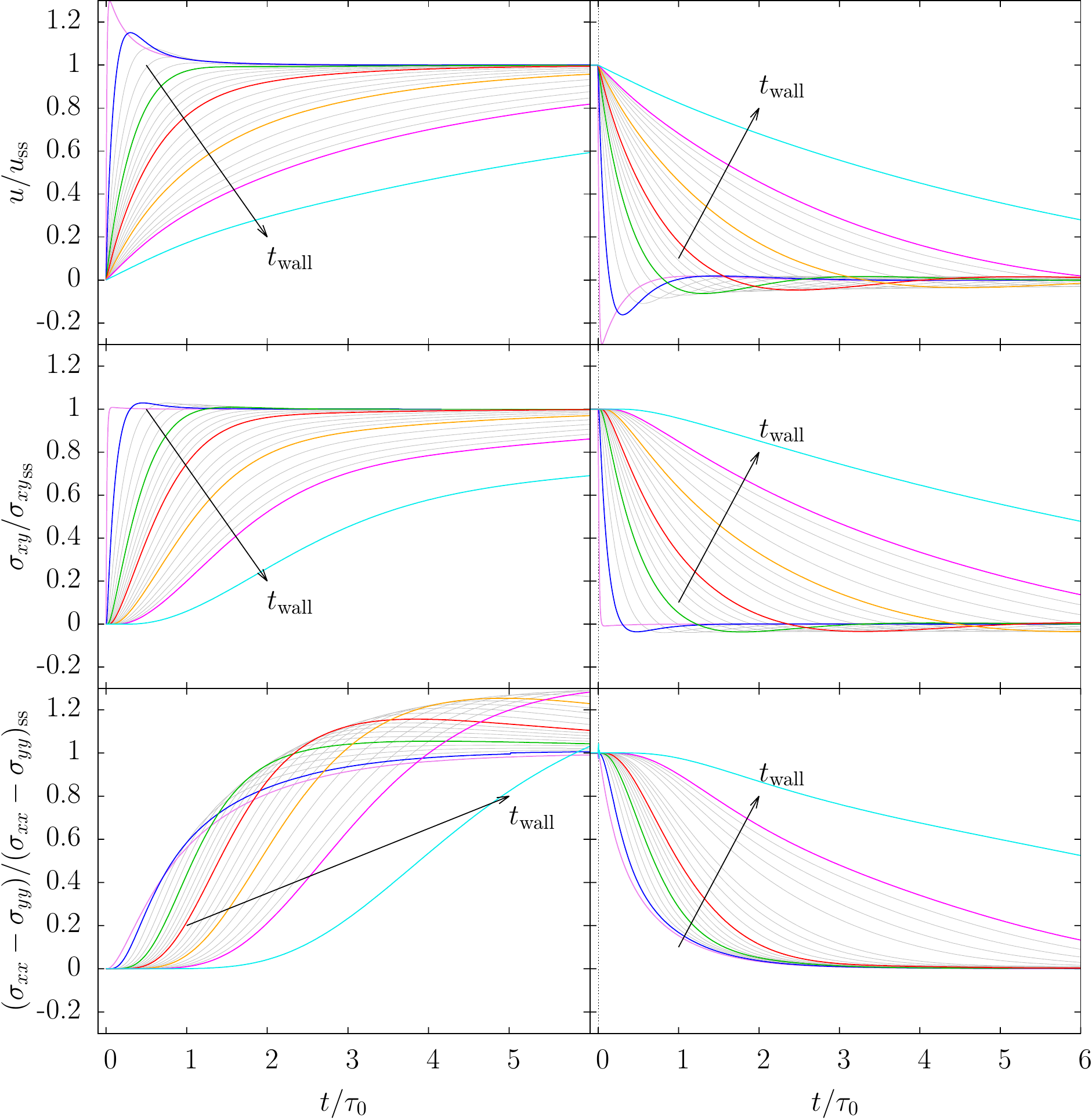}
\caption{\label{Qall}Velocity $u$ (top), shear stress $\sigma_{xy}$ (middle),
  and normal-stress difference $\sigma_{xx}-\sigma_{yy}$ (bottom) evolution
  after startup (left) and cessation (right) of pressure-driven channel
  flow of a nonlinear Maxwell fluid (relaxation time $\theta=10$),
  evaluated at position $d_{1/2}=y/H=0.51$ in the channel of width $2H$,
  and normalized by the respective steady-state values.
  Curves are shown for different channel widths corresponding to
  $\twall/\tau_0=0.01$, $0.1$, $0.5$, $1$, $2$, $4$, and $8$ (heavy lines);
  light grey lines mark values of $\twall$ equidistantly spaced in between.
}
\end{figure}

We now consider the transient dynamics of the channel flow upon applying
and removing a sudden driving pressure step.
Figure~\ref{Qall} presents an overview of the transient evolution for both
cases, for the velocity (top), shear-stress (middle), and normal-stress
difference (bottom) at a quarter-width of the channel (position
$d_{1/2}$ marked in Fig.~\ref{steadystate}. Curves are shown for
several channel widths. While for wide channels, the velocity $u(t)$
increases monotonically after application of the pressure gradient,
the evolution to the steady state is nonmonotonic for narrow channels.

In a Newtonian fluid, the channel diameter defines the
only relevant time scale of the transient dynamics. As long as the flow is
purely laminar, transverse-momentum diffusion across the channel sets the
time scale over which information of the boundary conditions is transmitted
to the center. This time scale is inversely proportional to the
Newtonian viscosity $\eta_\infty=G_\infty\tau_0$, and reads
\cite{Batchelor}
\begin{align}
\twall = \frac {4 H^2 \rho}{\pi^2 G_\infty \tau_0}.
\end{align}
As seen in Fig.~\ref{Qall}, $\twall\approx1$ separates the regime of narrow
channels from the regime of wide channels for the evolution of the startup
velocity. This can be retionalized as follows:
At short times after startup, $t\ll\tau_0$,
the fluid still behaves as a Newtonian fluid
with viscosity $\eta_\infty$, since the structural-relaxation contribution
to the stresses given by Eq.~\eqref{gitt} is still negligible.
For $\twall\ll\tau_0$, this time is sufficient to transiently build up the
parabolic velocity profile of the Newtonian fluid in the channel. Only at
times $t\gg\twall$, the flattened ``plug-like'' velocity profile of the
non-Newtonian fluid will be established. Hence, the transient velocity
first increases towards the larger Newtonian steady-state value, before it
decreases again to the lower non-Newtonian one. If $\twall\gg\tau_0$,
the initial high-shear Newtonian profile is not established, and
the velocity monotonically increases to the non-Newtonian steady state.

For the evolution of the shear stress $\sigma_{xy}(t)$
(middle left panel of Fig.~\ref{Qall}), $\twall$ marks the crossover
between the initial rise and a slower approach to the steady state.
The normal-stress differences shown in the bottom panel of the figure
display a more complex evolution towards the steady state, with overshoots
discernible for $\twall\gtrsim\tau_0$, and a monotonic increase for the
narrowest channels shown. This pattern depends on the position $y/H$
across the channel, as will be discussed below.

Starting and stopping flows are symmetric for Newtonian fluids, in the
sense that $u(t)$ after startup and $\tilde u(t)=u_\text{ss}-u(t)$ after
cessation are identical (where $u_\text{ss}$ is the steady-state value)
\cite{Batchelor}. The same symmetry holds for linear viscoelastic models
such as the UCM \cite{waters70}. It is broken for nonlinear constitutive
equations, since there the stress--strain-rate relations after cessation
(with no flow present) differ from that after startup (with flow present).
This can be seen in the right panels of Fig.~\ref{Qall}: the velocity
after cessation decays nonmonotonically towards zero even for
$\twall\gtrsim1$. The asymmetry is particularly clear for the
normal-stress difference. While its evolution after startup displays
overshoots for large $\twall$, the decays after cessation is always
monotonic, dictated by the fact that $\sigma_{xx}-\sigma_{yy}>0$ always
holds in the model for arbitrary time-dependent laminar channel flow.

\begin{figure}
\includegraphics[width=\linewidth]{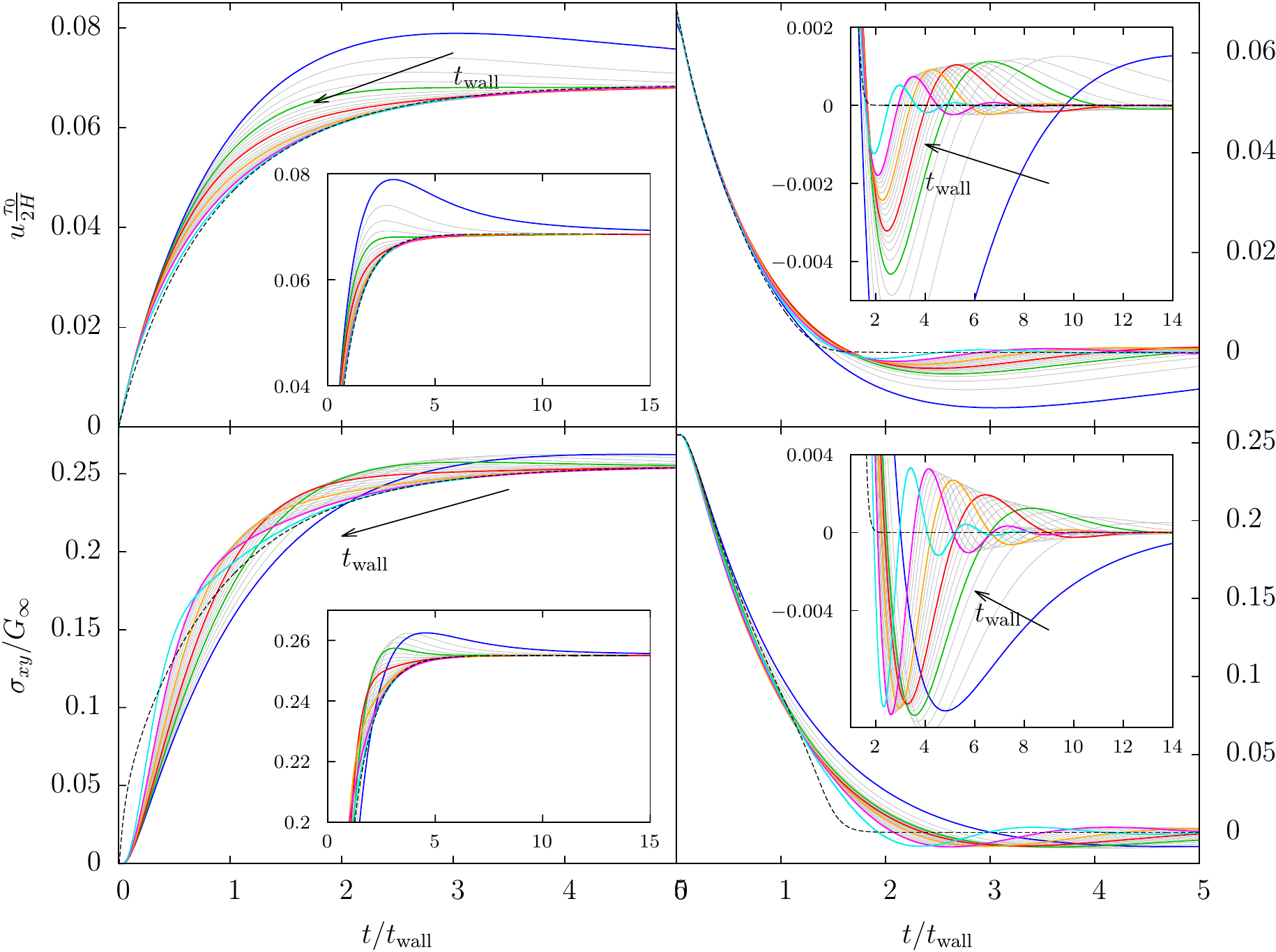}
\caption{\label{QT}Startup (left) and stopping (right) flow for the
  nonlinear Maxwell model for different channel diameters, as in
  Fig.~\ref{Qall}. The velocity $u(t)$ (top) and shear stress $\sigma_{xy}(t)$
  (bottom) are shown as functions of $t/\twall$. Dashed lines
  represent the instantaneous Maxwell model,
  cf.\ text.
}
\end{figure}

To demonstrate the relevance of $\twall$ for the initial startup and
cessation evolution, we show in Fig.~\ref{QT} the velocity and shear-stress
transients presented in Fig.~\ref{Qall} as functions of $t/\twall$.
For a Newtonian fluid, all curves for different channel widths would
collapse. The same holds for the instantaneous nonlinear Maxwell model,
Eqs.~\eqref{inlM}, discussed in Ref.~\onlinecite{LBpaper}. These results
are shown as dashed lines in Fig.~\ref{QT} for comparison.
Curves for the full nonlinear Maxwell model do not collapse, since
structural relaxation introduces a separate time scale that is independent
on $\twall$.

As shown in the right panels of Fig.~\ref{QT}, the decay of both the
velocity and the shear stress towards zero after removal of the pressure
gradient is oscillatory. The oscillations in the two quantities are
shifted in phase, following an initially faster decay of the velocity.
This is a consequence of the history integral appearing in Eq.~\eqref{gitt}:
as the velocity and with it the velocity gradients decay, the integral
determining the stresses is still dominated by past contributions.
Eventually, the velocity decays to zero, while stresses are still
present. To relax these stresses, the fluid continues to flow, but in a
direction opposite to the initial steady state (i.e., with negative velocity
since it is now driven by
the remaining internal stresses rather than the external pressure gradient).
This counter-flow causes the stresses to relax and eventually become negative,
such that the velocity starts to increase towards positive values again.

\begin{figure}
\includegraphics[width=\linewidth]{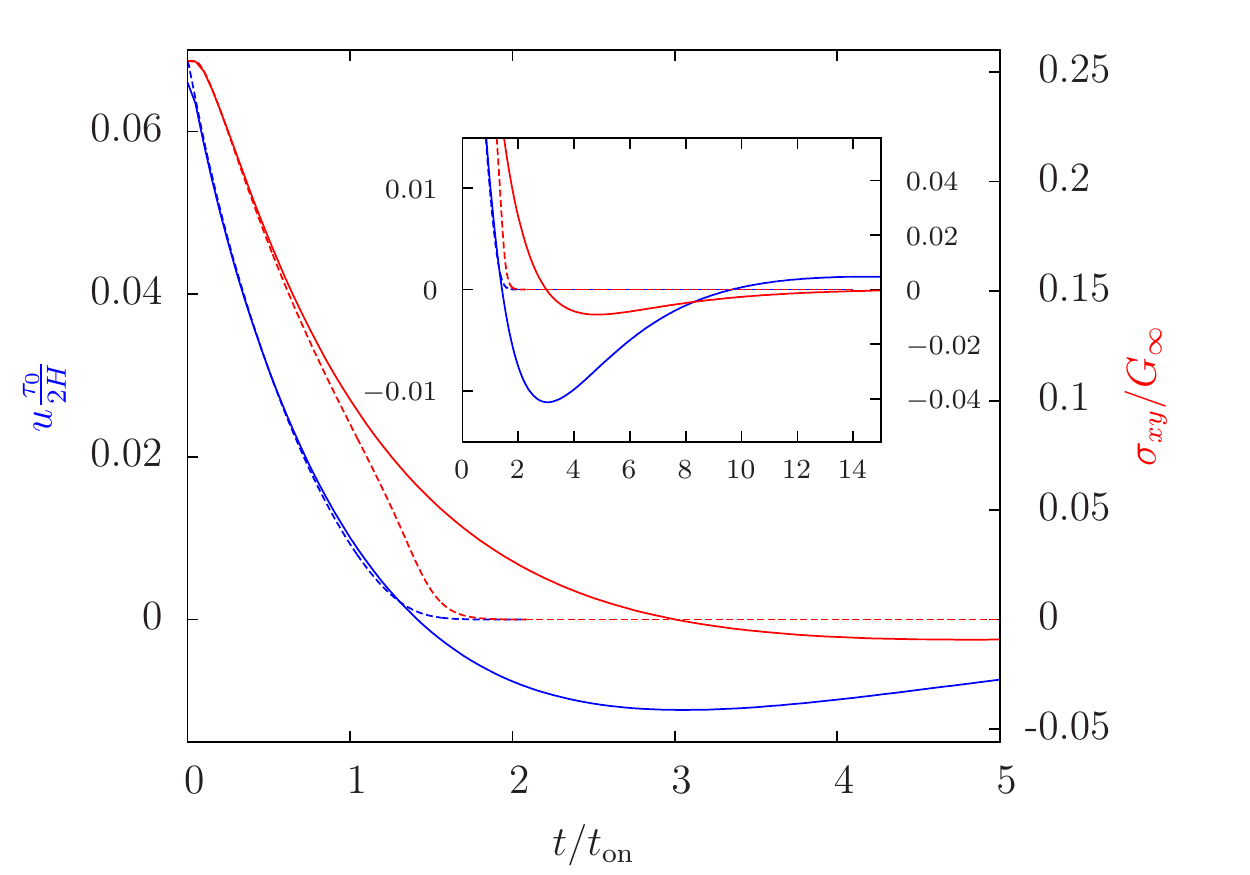}
\caption{\label{us}Evolution of velocity (left axis) and shear stress
  (right axis) as a function of $t/\twall$ after cessation of pressure
  driven channel flow, for a channel corresponding to $\twall=???$,
  at position $y/H=0.51$.
  Solid lines are results for the nonlinear Maxwell model,
  dashed lines represent the instantaneous model.
}
\end{figure}

Figure~\ref{us} highlights this phase-shifted oscillatory decay of the velocity
and the shear stress for a single channel width. It is instructive
to compare the observed decay pattern to that predicted by the instantaneous
model (dashed lines in Fig.~\ref{us}). Here, no oscillations are observed.
Up to the first zero crossing of the velocity seen for the integral model,
the decay of the velocity and the stress are very similar in the instantaneous
model. There, however, the stresses quickly decay to zero once the velocity
and its gradients vanish. The oscillatory behavior seen in the integral
model is hence a true signature of viscoelasticity.

For instantaneous models describing yield-stress fluids,
such as Eqs.~\eqref{inlM} in the limit $\tau\to\infty$, a finite stopping
time is observed in the decay of the velocity
\cite{Huilgol.2002,Huilgol.2002b,Chatzimina.2005}:
at some time $t_c$, the stresses in the channel have all decayed to values
below the yield stress, and hence the velocity has to obey $u(t)=0$
exactly for all $t>t_c$. Our model with $\tau=10\tau_0$ does not have a true
yield stress, but nevertheless the instantaneous model
shows the signature of this finite-time singularity. Increasing $\tau$,
the kink visible in the velocity decay around $t/\twall=1.2$ becomes more
pronounced. Realistic yield-stress fluids will typically be
viscoelastic, since the emergence of a yield stress is usually coupled
to slow structural relaxation and its modification through the flow.
In these fluids, the finite-time singularity $t_c$ does not mark the
exact coming to rest of the flow, but rather sets a typical time scale
for the oscillatory decay of velocities and stresses.

\begin{figure}
\includegraphics[width=\linewidth]{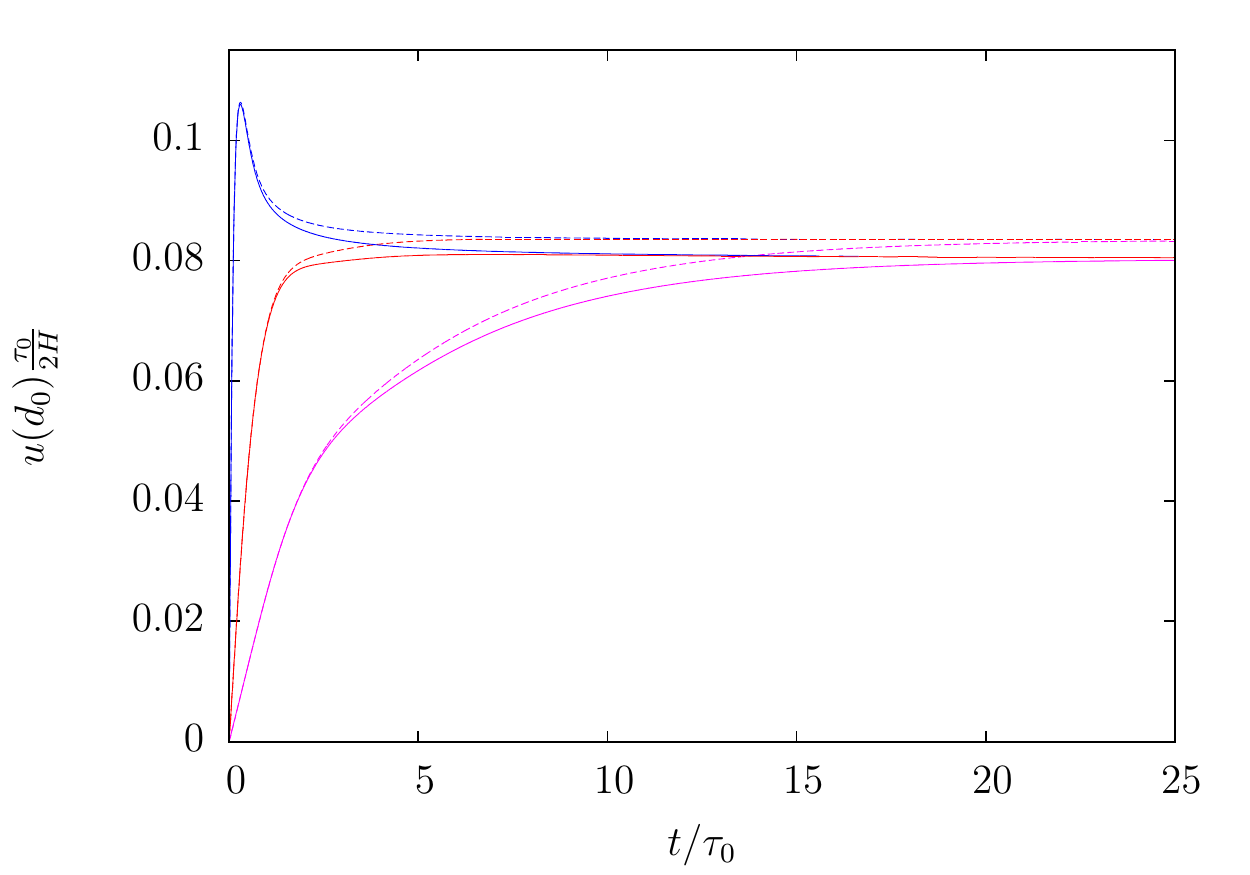}
\caption{\label{figtau}Startup velocity as in Fig.~\ref{Qall}, for channel
  widths corresponding to $\twall/\tau_0=0.1$, $1$, and $4$, in the
  integral nonlinear Maxwell model with different structural relaxation
  times ($\theta=10$: dashed; $\theta=100$: solid), evaluated at
  position $y/H=0.51$.
}
\end{figure}

We briefly discuss the influence of the structural relaxation time on the
transients. Figure~\ref{figtau} compares the startup velocities for
selected $\twall$ (also shown in Fig.~\ref{Qall}) for $\tau=10\tau_0$
and $\tau=100\tau_0$. These curves differ essentially only by the
different steady-state values the tend to. For the smaller $\tau$,
the steady-state velocities are higher, as the fluid has a lower viscosity
in the low-shear region near the center of the channel. Still, for
$\twall\ll\tau_0$, an overshoot is seen that vanishes for $\twall\gg\tau_0$.
Note that in our definition of $\twall$, the structural relaxation time
$\tau$ does not enter. This may appear surprising, since for viscoelastic
models, one might expect a strong dependence of the transient dynamics
on the structural relaxation time.

\begin{figure}
\includegraphics[width=\linewidth]{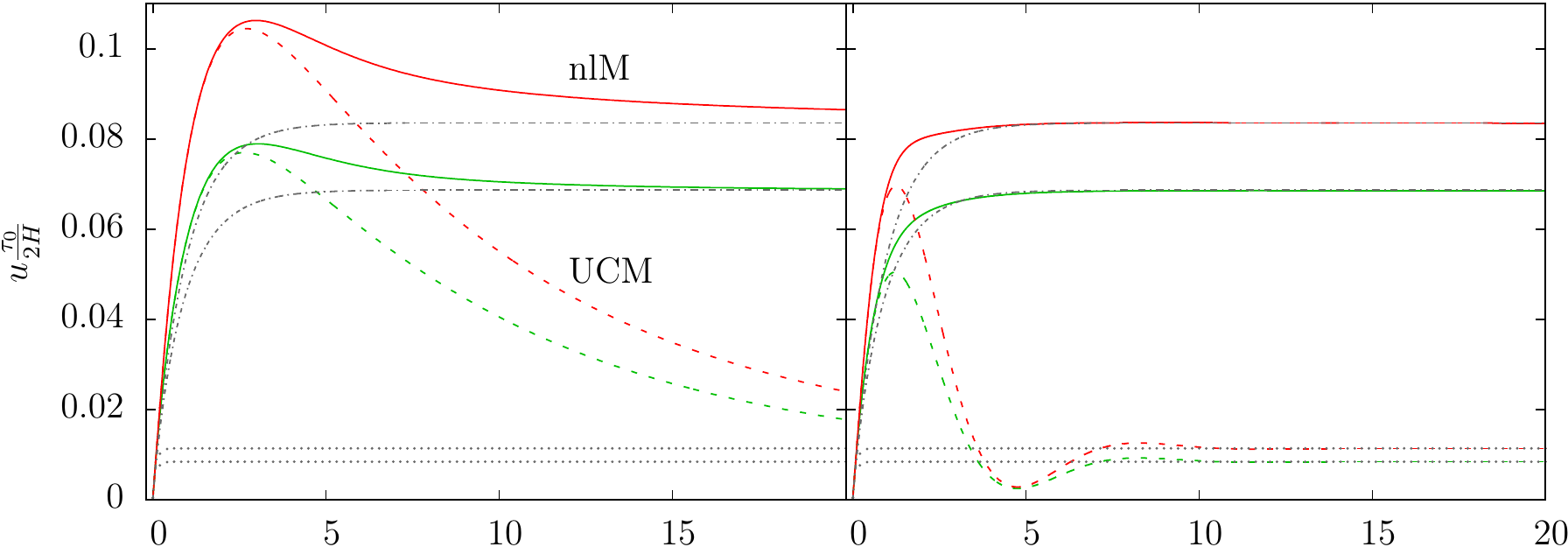}
\caption{\label{compModels}Starting flow of the velocity
  in a narrow channel (left, $\twall/\tau_0=0.1$) and an intermediate-width
  channel (right, $\twall/\tau_0=1$), evaluated at positions $y/H=d_0$
  and $d_{1/2}$ marked in Fig.~\ref{steadystate}. Solid lines represent
  the nonlinear Maxwell model including shear thinning,
  dashed lines the linear-viscoelastic UCM model, both for $\theta=10$.
  Dash-dotted and dotted lines represent the instantaneous models
  with and without shear thinning (Eq.~\eqref{inlM} and a Newtonian fluid,
  respectively).
}
\end{figure} 

To highlight the separate effects of (linear) viscoelasticity and
shear thinning, we compare in Fig.~\ref{compModels} the startup velocities
for the integral nonlinear Maxwell model with those for the purely
linear-viscoelastic UCM for. The latter does not include shear thinning,
so that the steady-state velocities are much lower, owing to the high
viscosity set by the structural relaxation time $\tau$. The transients
observed in the UCM display overshoots (and, in fact, oscillations)
for all channel widths, as highlighted
for both $\twall\ll\tau_0$ and $\twall=\tau_0$ in the figure.
The inclusion of shear thinning into this viscoelastic model causes the
oscillations to disappear. Realistic viscoelastic fluids will most
likely also show shear thinning, since the appearance of slow structural
relaxation makes the system prone to exhibit nonlinear-response phenomena.
The qualitative behavior of the transient flow dynamics should hence
be closer to our nonlinear model than to the UCM.

\subsection{Profile Evolution}

So far, we have discussed the time evolution of velocities and stresses
at selected positions across the channel. For a Newtonian fluid or
the instantaneous model \cite{LBpaper}, this contains the essential
information, since the shape of the cross-channel profiles does not
change qualitatively during startup or cessation of the flow.

\begin{figure}
\includegraphics[width=\linewidth]{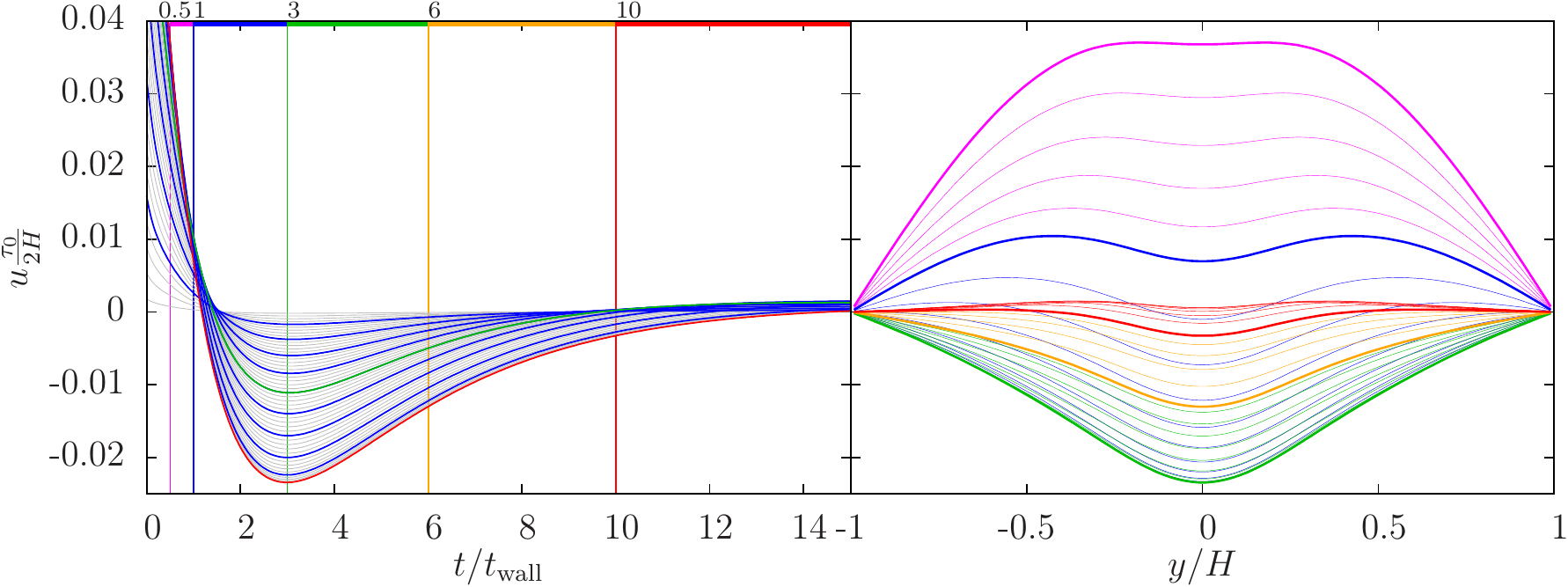}
\includegraphics[width=\linewidth]{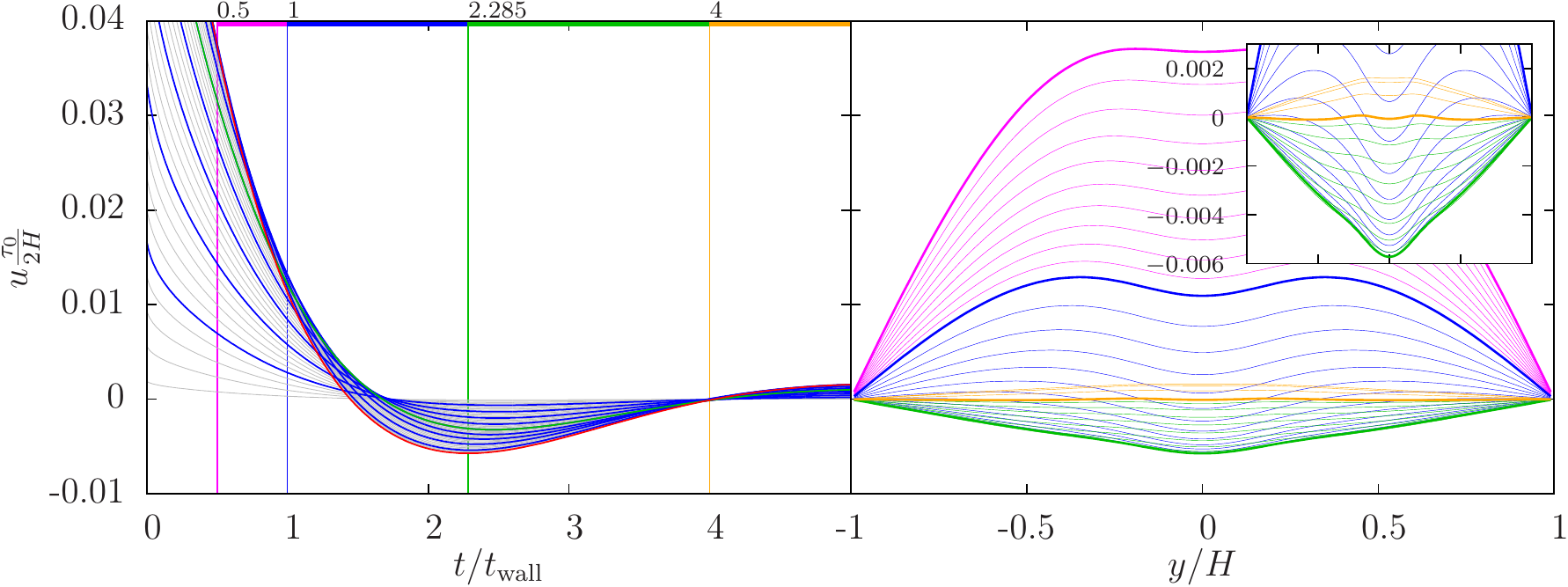}
\caption{\label{uoffcut}Stopping flow of the velocity for the narrow
  (top, $\twall/\tau_0=0.1$) and intermediate (bottom,
  $\twall/\tau_0=1$) channel, for the integral nonlinear Maxwell model.
  The profiles (right) are plotted for
  different times in intervals indicated by a horizontal line of the same
  color at the top in the left panels.
  Each line of the same color is separated by $\Delta
  t$, which is doubled with each new color starting with $\Delta
  t^\text{magenta}=0.1\twall$ and $0.05\twall$, respectively. Profiles
  plotted with bold lines are taken at times marked by vertical
  lines in the left panels.}
\end{figure}

The profile evolution in particular during cessatino of the flow is
more complex for the present integral nonlinear Maxwell model. As
shown in Fig.~\ref{uoffcut}, the velocity first relaxes to zero at
slightly different times, depending on the cross-channel position.
Interestingly, the plug in the center of the channel does not come
to rest as a plug, but rather the velocity around $y/H=0$ decreases
faster than the nearby velocities. For the UCM without shear thinning,
but including the Newtonian high-shear viscosity $\eta_\infty$,
this is not observed, and rather the center-channel velocity is slower
to decay.

The decay pattern of the velocities transmits to an intricate decay
pattern also for the normal-stress difference. Recall that we are
considering laminar, translation-invariant, incompressible flow.
The normal stresses
are hence determined by the velocity gradients, but do not couple back
to the flow evolution.

\begin{figure}
\includegraphics[width=\linewidth]{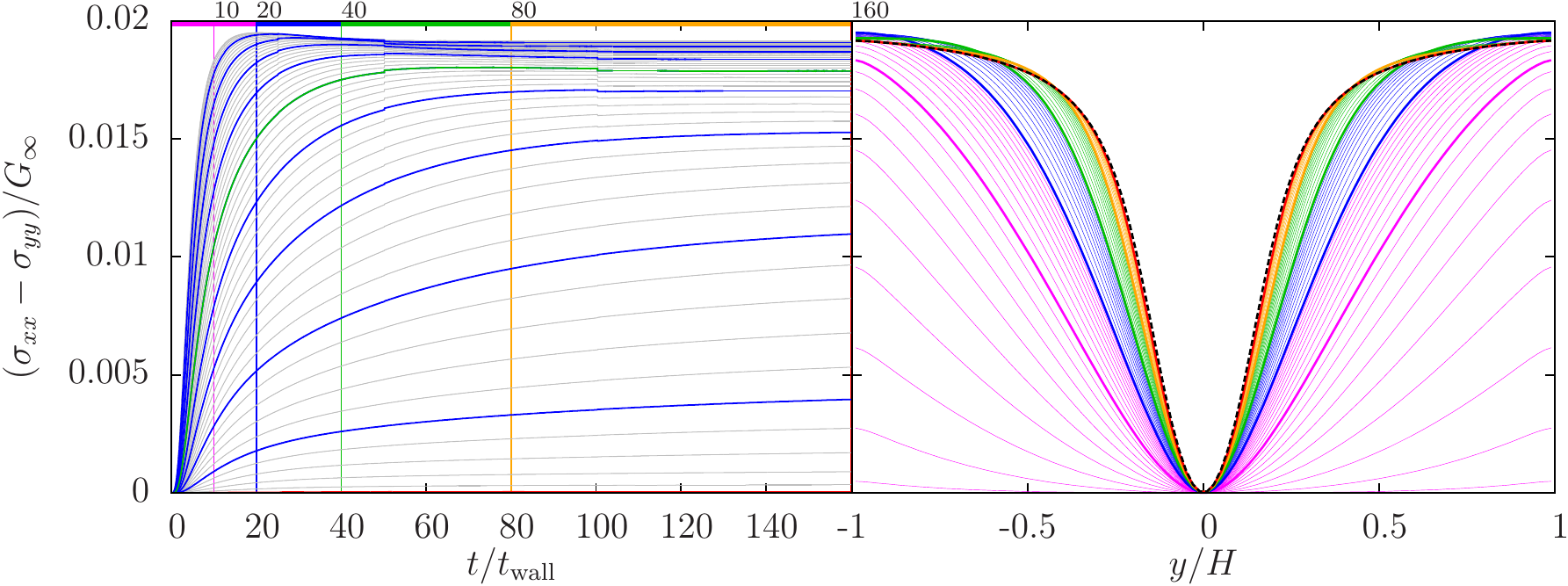}
\includegraphics[width=\linewidth]{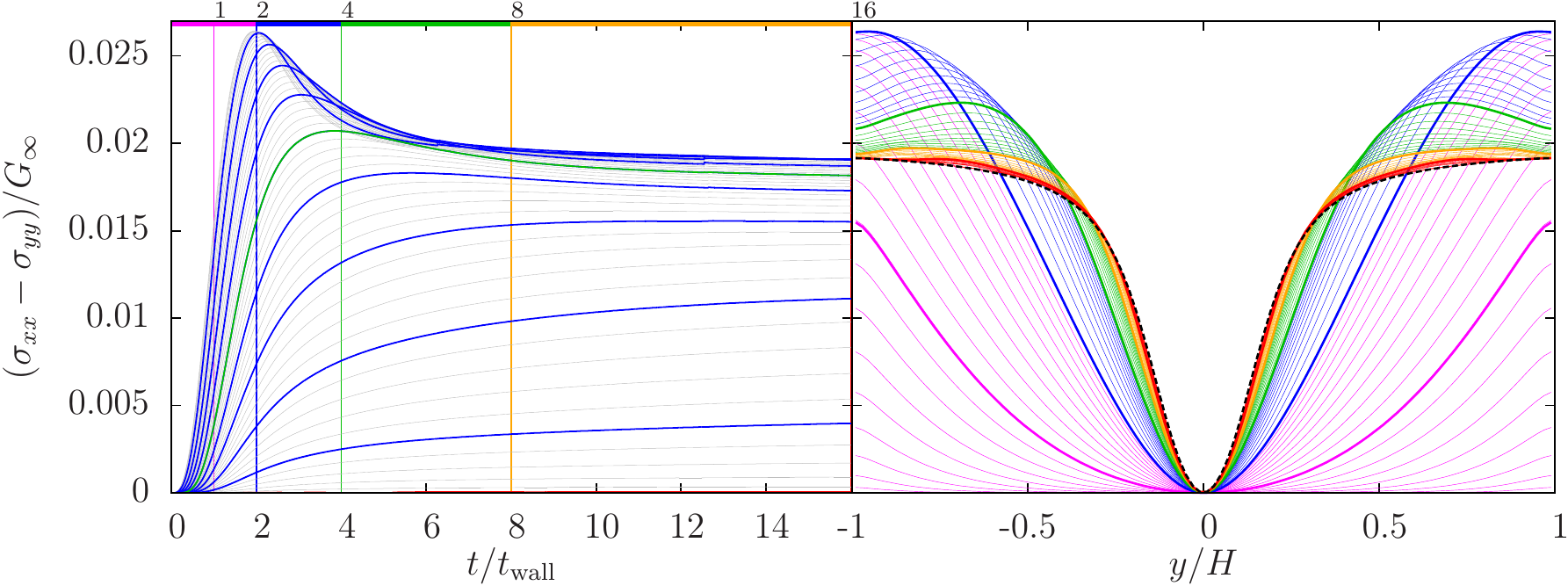}
\caption{\label{ncut}Startup flow of the normal stress difference for
  the narrow (top) and intermediate (bottom) channel. The profiles
  (right) are plotted for different times indicated by a horizontal
  line of the same color at the top (left). Each line of the same
  color is separated by $\Delta t$, which is doubled with each new
  color starting with $\Delta t^\text{magenta}=0.1\twall$ and $1
  \twall$, respectively. Profiles plotted with bold lines are
  taken at times marked by vertical lines, the dashed line is the
  steady state profile.
}
\end{figure}

Figure~\ref{ncut} shows the evolution of the normal-stress-difference
profiles in startup flow, for the two channel widths discussed
also in Fig.~\ref{uoffcut}. These profiles evolve towards the
characteristic steady-state profile which is quadratic in the center
of the channel, crossing over to a constant near the walls. In the
wider channel ($\twall=\tau_0$), the increase in normal-stress difference
close to the wall is more pronounced, and causes transient profiles that
display a minimum at $y/H=0$, and a maximum between $y/H=0.5$ and $1$.
In the cuts at constant $y/H$ discussed above (see also left panel
of the figure), this manifests itself as an overshoot in the
time evolution which is not present for the narrower channel. Hence
this overshoot is dominated by effects close to the channel wall, where
high shear rates transiently cause large normal-stress differences. Note
that small compressibility effects and normal-stress-induced particle migration
might change this behavior.

\begin{figure}
\includegraphics[width=\linewidth]{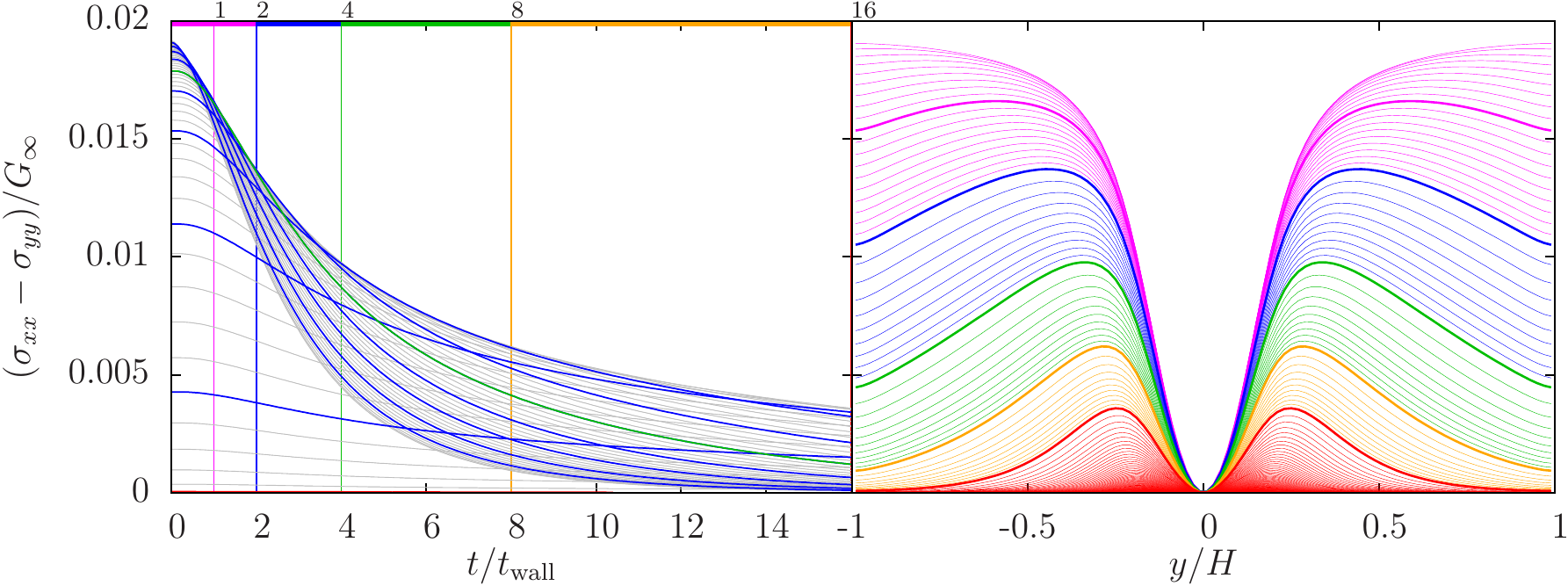}
\includegraphics[width=\linewidth]{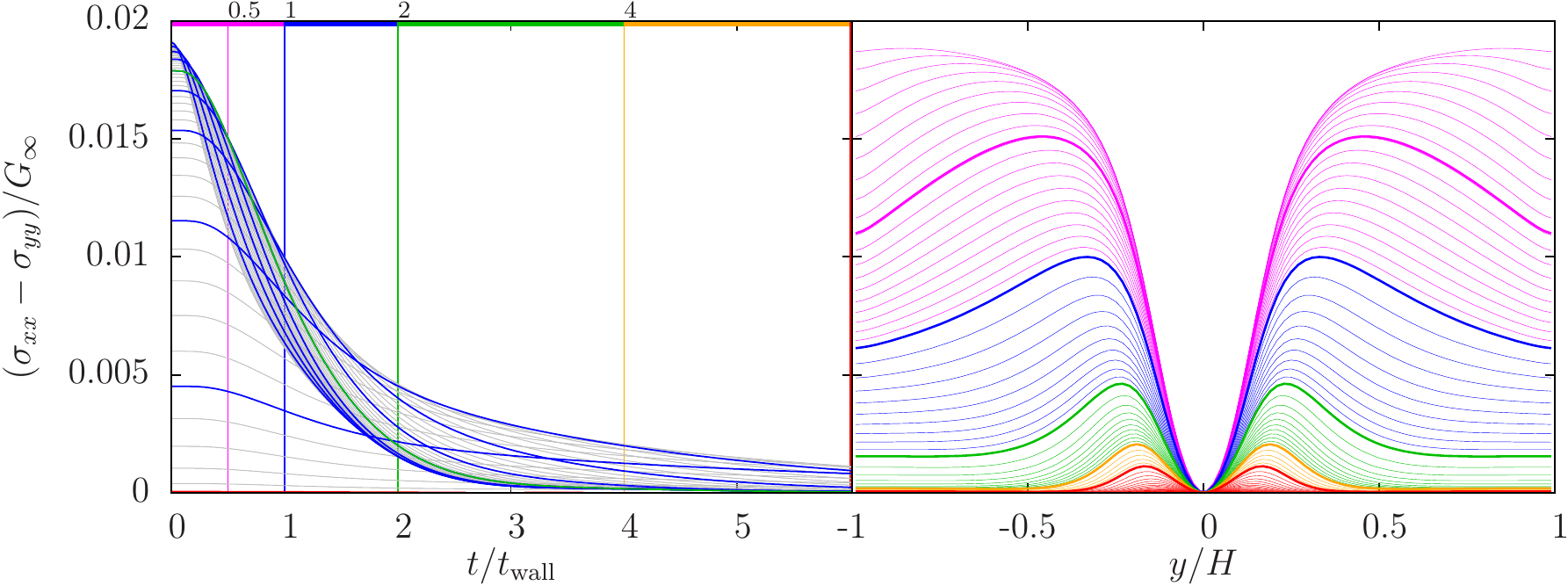}
\caption{\label{noffcut}Stopping flow of the normal stress difference
  for the narrow (top) and intermediate (bottom) channel. The profiles
  (right) are plotted for different times indicated by a horizontal
  line of the same color at the top (left). Each line of the same
  color is separated by $\Delta t$, which is doubled with each new
  color starting with $\Delta t^\text{magenta}=0.1\twall$ and
  $0.05\twall$, respectively. Profiles plotted with bold lines
  are taken at times marked by vertical lines.}
\end{figure}

As mentioned above, in the cessation flow, normal-stress differences decay
monotonically since they have to remain positive at all times. The
corresponding profiles are shown in Fig.~\ref{noffcut}.
Again, the normal-stress difference shows the fastes transient evolution
near the walls.
Since in the center of the channel, the normal-stress difference is close
to zero even in steady-state, this results in profiles that again have
a minimum around $y/H=0$ and a maximum at intermediate $y/H$. Different
from what is seen in Fig.~\ref{ncut} for the startup flow, in cessation,
no qualitative change is observed between the narrow and a wider channel.

\section{Conclusion}\label{sec:concl}

We have developed a hybrid-lattice-Boltzmann simulation scheme that allows
to simulate the flow of non-Newtonian, glass-forming fluids incorporating
flow-history effects that arise from slow structural relaxation. The
scheme builds upon an extension of the standard LB scheme to non-Newtonian
constitutive equations presented in Ref.~\cite{LBpaper}. Here we extend
it to include an integral-equation solver adapted to constitutive
equations of the form, Eq.~\eqref{gitt}, generically expected in nonlinear
glassy rheology. The scheme is particularly adapted to deal with
flows that include long-lived memory effects.

The hybrid-LB algorithm was used to study the combined effects of
viscoelasticity and shear thinning in pressure-driven planar channel
flow of an incompressible fluid. To mimic features expected from microscopic
theory, such as ITT-MCT \cite{brader09}, we have employed a nonlinear
generalized Maxwell model.
The steady-state profiles of this model have been discussed earlier.
As is typical for a fluid close to the glass transition, plug-like
flow develops in the center of the channel, as a signature of the
yield stress that arises at the glass transition.

The transient evolution of velocities, shear stresses, and normal-stress
differences is rich in phenomenology. Since the time scale $\twall$
characterizing hydrodynamic momentum transport across the channel
is independent on the structural relaxation phenomena, the transient
evolution of the flow differs in narrow channels from that in wide channels.
For channel widths corresponding to $\twall\ll\tau_0$, overshoots appear
in the startup velocity. Note that
$\twall\approx\tau_0$ corresponds to channel widths $H=O(\text{mm})$,
so that the effects should readily be observable in experiment, e.g.,
on colloidal suspensions.

The decay of the velocities and stresses after removal of the driving
pressure gradient is oscillatory, reflecting the viscoelasticity that
causes the stress evolution to lag behind that of the velocity.
Transiently, the system comes to a rest, at a time given by the
finite-time singularity discussed for yield-stress fluids with an
instantaneous relation between stress and strain rate. As at this time,
the stress has not fully decayed, it causes a backward motion of the fluid,
and hence slower oscillatory approach to rest. Such oscillations are
already expected from linear viscoelasticity, as exemplified by the
upper-convected Maxwell model. However, generically, glass-forming
fluids will exhibit both viscoelasticity and shear thinning.

The nonlinear Maxwell model implemented here should account for many
qualitative effects expected from more microscopic theories, such as
ITT-MCT. Our hybrid-LB algorithm is readily adapted to include
constitutive equations directly taken from ITT-MCT, although they are
numerically much more demanding, since the ad-hoc exponential assumed
in Eq.~\eqref{nlmax} is replaced by an expression evolving density-correlation
functions that need to be calculated from the solution of integro-differential
equations with long-lived memory kernels. Such a combined MCT-LB scheme
will be the subject of future work.

\begin{acknowledgments}
This work was supported by DFG Research Unit FOR1394, project P3.
\end{acknowledgments}

\begin{appendix}
\section{Upper-convected Maxwell model}
\label{sec:appendix}

\begin{figure}
\includegraphics[width=\linewidth]{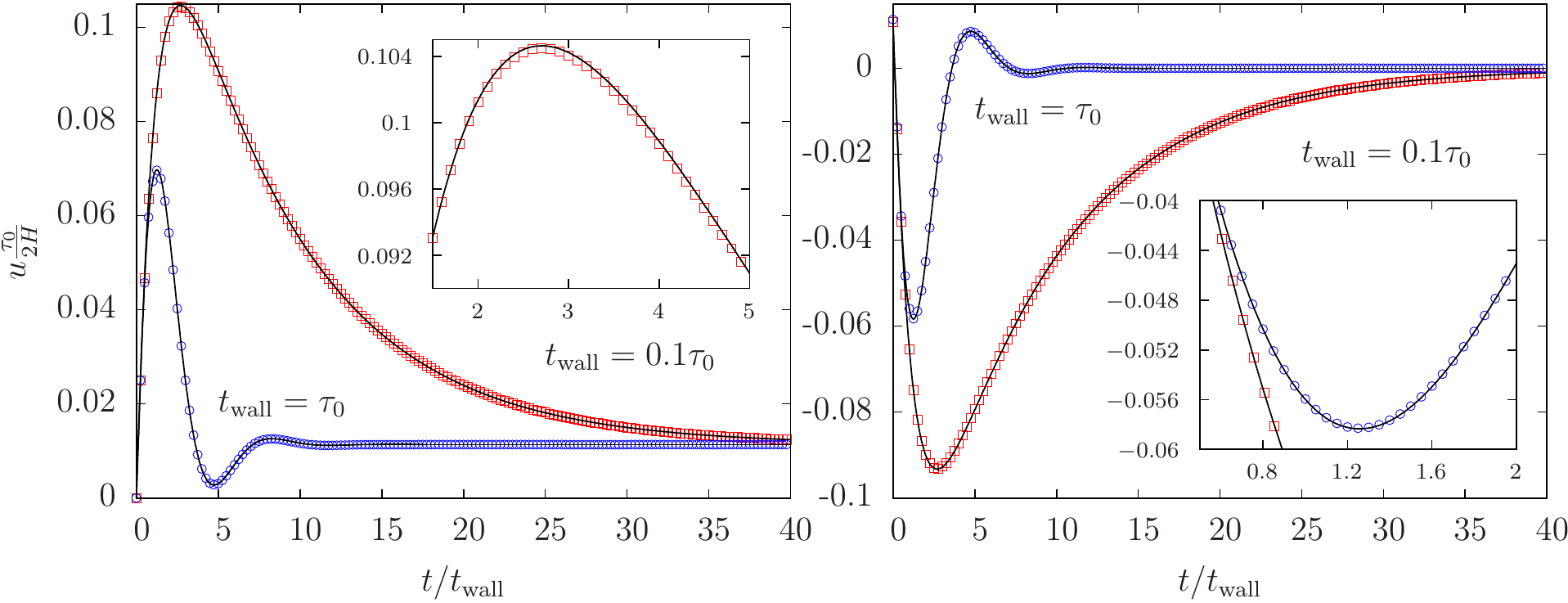}
\caption{\label{UCM}Startup (left) and stopping flow (right) for the
  UCM model for two channel diameters with $\twall=0.1\tau_0$ (squares)
  and $\twall=\tau_0$ (circles). The velocity is measured at $d=0.51$,
  lines are analytic results and symbols are obtained from LB
  simulations.}
\end{figure}
The 2D Poiseuille flow of a UCM fluid has been solved
analytically\cite{waters70} and provides a good test for our scheme to
implement viscoelasticity in lattice Boltzmann simulations via an
integral constitutive equation. Fig.~\ref{UCM} shows the center
velocity ($d=y/H=0.51$) for two different channel diameters with
$\twall=0.1\tau_0$ (red) and $\twall=\tau_0$ (blue) after applying
(left) and removing (right) a sudden pressure difference. The channel
diameters are the ones we find most interesting when discussing the
nonlinear Maxwell model. Although only $100$ nodes in transverse flow
direction were used, the LB simulations reproduce the analytic results
(black lines) extremely well. Deviations are larger in amplitude than
in time and a higher precision can be easily attained by increasing
the lattice size. The algorithm shows the same precision for the
stopping flow as under startup and reproduces the symmetry.

\begin{figure}
\includegraphics[width=\linewidth]{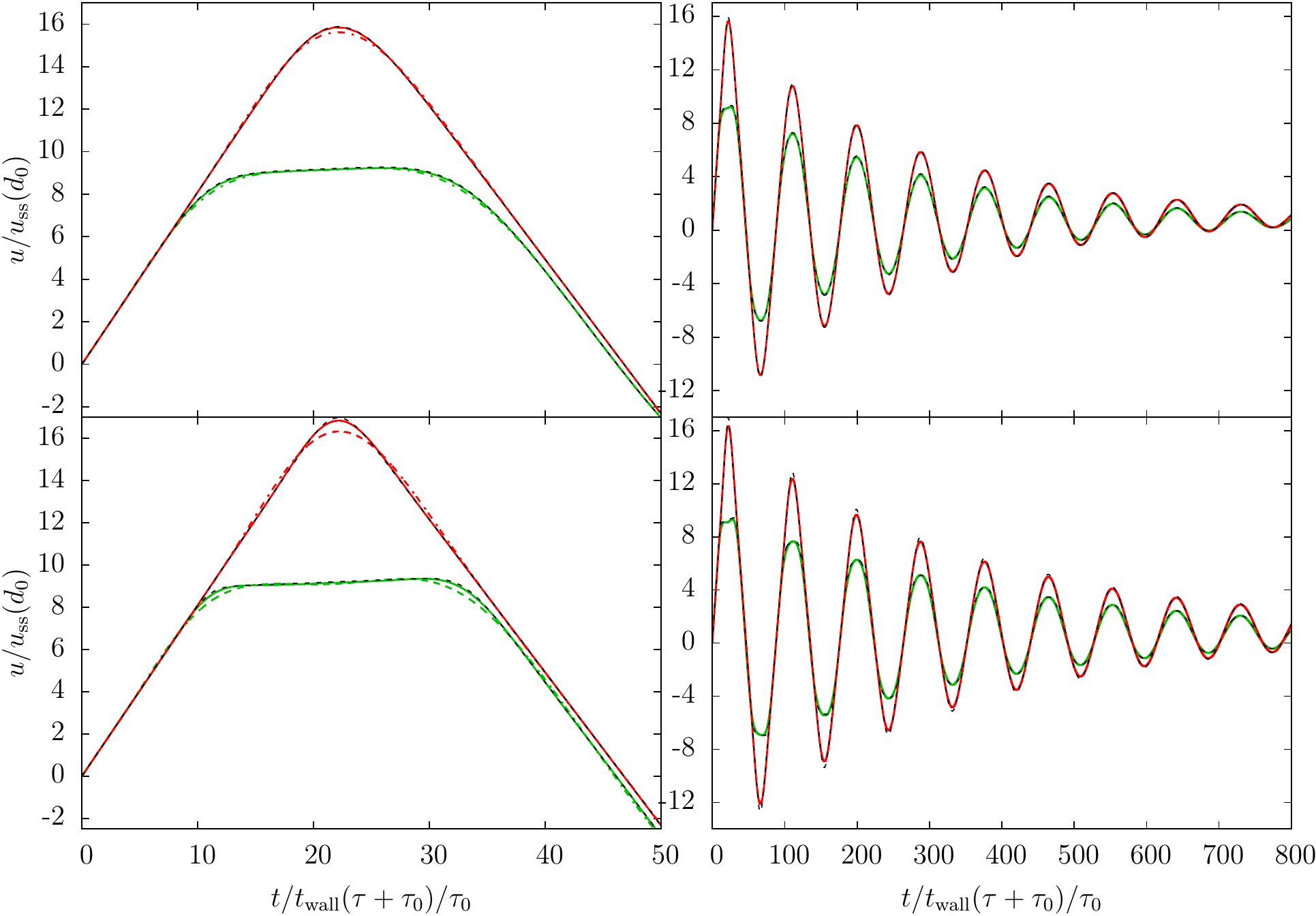}
\caption{\label{funnyMaxwell}Velocity profile of the UCM model with
  $\theta=400$, $L=\sqrt 8\unit{m}$ (top) and $\theta=2000$,
  $\sqrt{200}\unit{m}$ (bottom) using lattices with $N=800$ (solid
  lines) and $N=200$ nodes (dash-dotted lines), and the analytical
  solution (dashed lines). The velocity is measured in the center
  $d_0=\{0.005, 0.00125\}$ (red) and at quarter position $d_{\sfrac 1 2}=\{0.505,
  0.50125\}$ (green) for $N=\{200, 800\}$.
}
\end{figure}

To further test the capabilities of the LB scheme, we consider highly
viscous UCM fluids in very wide channels, $\theta=\{400, 2000\}$ and
$2H=L=\{\sqrt{8}, \sqrt{200}\} \unit m$. This way, the dimensionless
retardation time\cite{waters70} $S_2=G_\infty(\tau+\tau_0)\tau_0/(\rho
L^2)=\{0.05, 0.01\}\ll 1$, but the relaxation time
$S_1=G_\infty(\tau+\tau_0)\tau/(\rho L^2)=\{20.05, 20.01\}\gg 1$. The
density is $\rho=1000 \unit{kg/m^3}$, the shear modulus $G_\infty=1
\unit{Pa}$. This limit is interesting as the UCM fluid behaves similar
to a soft elastic solid, but the large channel diameter allows the
material to deform for a long time unperturbed by boundary
effects. Fig~\ref{funnyMaxwell} shows the evolution of the center and
half-center velocity in time. Please note, that the time is now given
in units of $\twall\tau_0/(\tau+\tau_0)\propto \eta_\text{max}^{-1}$,
where $\eta_\text{max}=G_\infty(\tau+\tau_0)$ is the long time limit
of the viscosity under steady shear. Previously, the transient
dynamics was accelerated by shear and we identified $\twall\propto
\eta_\infty^{-1}$ as the characteristic transient time scale of the
nonlinear Maxwell model.

\begin{figure}
\includegraphics[width=\linewidth]{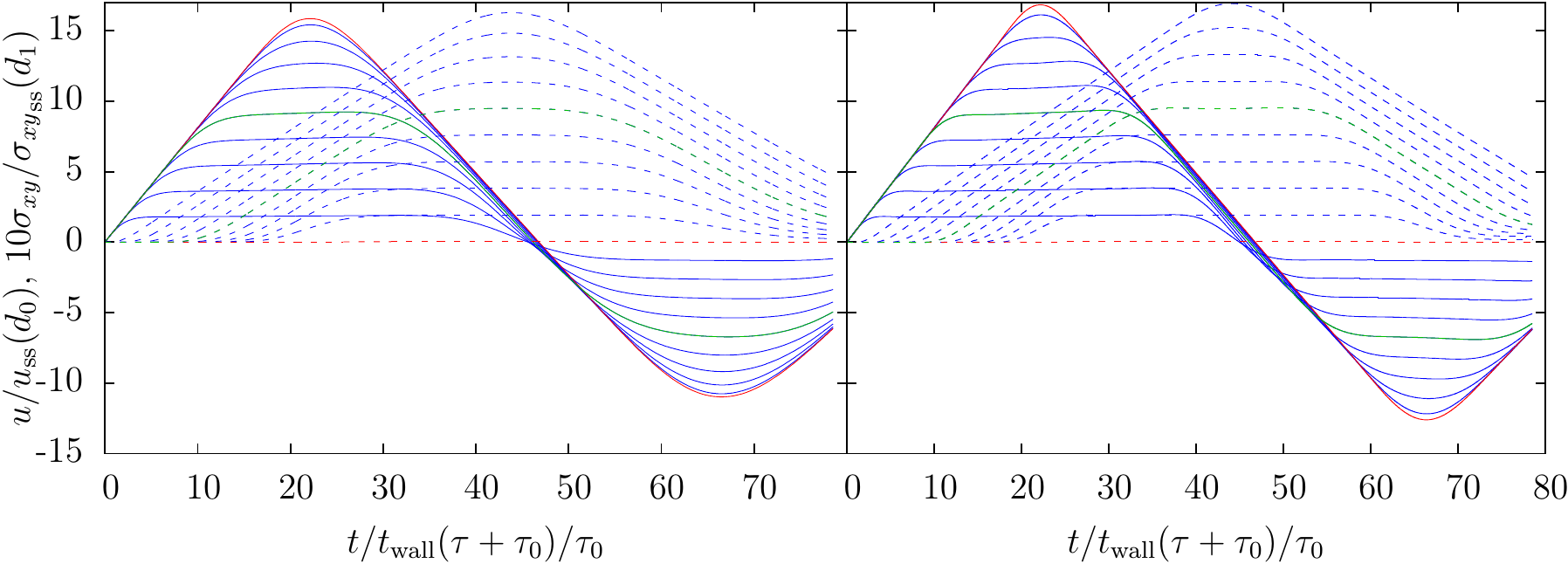}
\caption{\label{fMvelstr} UCM. Velocity (solid lines) and shear stress
  (dashed lines, multiplied by $10$) for $\theta=400$ (left) and
  $\theta=2000$ (right). The velocity and shear stress are scaled by
  their maximum steady state value.}
\end{figure}

The initial response of the UCM model is almost purely
elastic. Fig.~\ref{fMvelstr} shows the evolution of the velocity
(solid lines) and shear stress (dashed lines) in time at equally
spaced positions in the channel. Red lines are in the channel center,
green lines at half-center. The applied pressure gradient exerts a
constant body force on the fluid. As viscous damping takes place on a
time scale much larger than the elastic response, the velocity
increases linearly from each wall to develop a homogeneous shear
field. Once the shear waves meet in the channel center, the fluid
slows down again. The shear stress only starts to build up, when the
velocity gradient is almost constant. It then increases linearly to
satisfy the constant stress to strain relation of the Maxwell model.
The driving force is in turns used to increase either the kinetic
energy or stress of the fluid.

Viscous damping forces are small initially and only affect the
dynamics on long time scales. As $\theta$ is large, the memory of the
initial state only decays slowly to finally give a flowing steady
state much slower than the initial response after applying the
pressure gradient, see Fig.~\ref{funnyMaxwell}. The
lattice Boltzmann scheme is able to track this long time evolution
even for small lattice sizes. More precise results, especially for the
first period, can be obtained when a larger lattice is used. We find
the largest deviations from the analytic solution when there are sudden
changes in the velocity due to the solid-like dynamics, as the LB
algorithm always assumes a fluid.

\begin{figure}
\includegraphics[width=\linewidth]{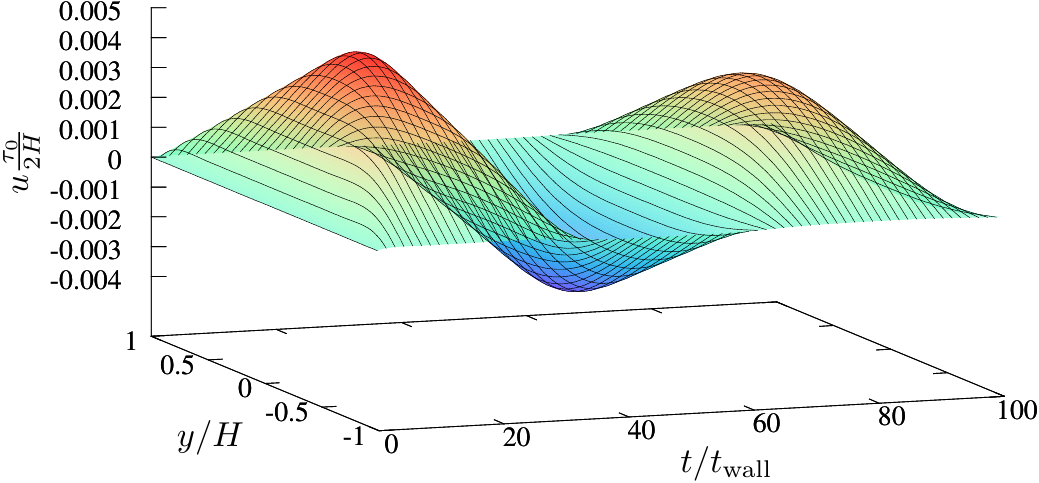} 
\caption{\label{3DUCM}UCM model. $S_1=20.05, S_2=0.05$. $\theta=400$, $L=\sqrt 8\unit{m}$. Bump in profile for small times.
}
\end{figure}

\vspace{2cm}
\end{appendix}

\bibliography{lit}

\begin{thebibliography}{42}%
\makeatletter
\providecommand \@ifxundefined [1]{%
 \@ifx{#1\undefined}
}%
\providecommand \@ifnum [1]{%
 \ifnum #1\expandafter \@firstoftwo
 \else \expandafter \@secondoftwo
 \fi
}%
\providecommand \@ifx [1]{%
 \ifx #1\expandafter \@firstoftwo
 \else \expandafter \@secondoftwo
 \fi
}%
\providecommand \natexlab [1]{#1}%
\providecommand \enquote  [1]{``#1''}%
\providecommand \bibnamefont  [1]{#1}%
\providecommand \bibfnamefont [1]{#1}%
\providecommand \citenamefont [1]{#1}%
\providecommand \href@noop [0]{\@secondoftwo}%
\providecommand \href [0]{\begingroup \@sanitize@url \@href}%
\providecommand \@href[1]{\@@startlink{#1}\@@href}%
\providecommand \@@href[1]{\endgroup#1\@@endlink}%
\providecommand \@sanitize@url [0]{\catcode `\\12\catcode `\$12\catcode
  `\&12\catcode `\#12\catcode `\^12\catcode `\_12\catcode `\%12\relax}%
\providecommand \@@startlink[1]{}%
\providecommand \@@endlink[0]{}%
\providecommand \url  [0]{\begingroup\@sanitize@url \@url }%
\providecommand \@url [1]{\endgroup\@href {#1}{\urlprefix }}%
\providecommand \urlprefix  [0]{URL }%
\providecommand \Eprint [0]{\href }%
\providecommand \doibase [0]{http://dx.doi.org/}%
\providecommand \selectlanguage [0]{\@gobble}%
\providecommand \bibinfo  [0]{\@secondoftwo}%
\providecommand \bibfield  [0]{\@secondoftwo}%
\providecommand \translation [1]{[#1]}%
\providecommand \BibitemOpen [0]{}%
\providecommand \bibitemStop [0]{}%
\providecommand \bibitemNoStop [0]{.\EOS\space}%
\providecommand \EOS [0]{\spacefactor3000\relax}%
\providecommand \BibitemShut  [1]{\csname bibitem#1\endcsname}%
\let\auto@bib@innerbib\@empty
\bibitem [{\citenamefont {Voigtmann}(2014)}]{Voigtmann14cocis}%
  \BibitemOpen
  \bibfield  {author} {\bibinfo {author} {\bibfnamefont {{\relax
  Th}.}~\bibnamefont {Voigtmann}},\ }\href@noop {} {\bibfield  {journal}
  {\bibinfo  {journal} {Curr. Opin. Colloid Interf. Sci.}\ }\textbf {\bibinfo
  {volume} {19}},\ \bibinfo {pages} {49} (\bibinfo {year} {2014})}\BibitemShut
  {NoStop}%
\bibitem [{\citenamefont {Maxwell}(1867)}]{Maxwell1867}%
  \BibitemOpen
  \bibfield  {author} {\bibinfo {author} {\bibfnamefont {J.~C.}\ \bibnamefont
  {Maxwell}},\ }\href@noop {} {\bibfield  {journal} {\bibinfo  {journal} {Phil.
  Trans. R. Soc. London}\ }\textbf {\bibinfo {volume} {157}},\ \bibinfo {pages}
  {49} (\bibinfo {year} {1867})}\BibitemShut {NoStop}%
\bibitem [{\citenamefont {Salen\c{c}on}(2001)}]{Salencon.2001}%
  \BibitemOpen
  \bibfield  {author} {\bibinfo {author} {\bibfnamefont {J.}~\bibnamefont
  {Salen\c{c}on}},\ }\href@noop {} {\emph {\bibinfo {title} {Handbook of
  Continuum Mechanics}}}\ (\bibinfo  {publisher} {Springer-Verlag},\ \bibinfo
  {address} {Berlin},\ \bibinfo {year} {2001})\BibitemShut {NoStop}%
\bibitem [{\citenamefont {Fuchs}\ and\ \citenamefont {Cates}(2002)}]{fuchs02}%
  \BibitemOpen
  \bibfield  {author} {\bibinfo {author} {\bibfnamefont {M.}~\bibnamefont
  {Fuchs}}\ and\ \bibinfo {author} {\bibfnamefont {M.~E.}\ \bibnamefont
  {Cates}},\ }\href {\doibase 10.1103/PhysRevLett.89.248304} {\bibfield
  {journal} {\bibinfo  {journal} {Phys. Rev. Lett.}\ }\textbf {\bibinfo
  {volume} {89}},\ \bibinfo {pages} {248304} (\bibinfo {year}
  {2002})}\BibitemShut {NoStop}%
\bibitem [{\citenamefont {Fuchs}\ and\ \citenamefont {Cates}(2009)}]{fuchs09}%
  \BibitemOpen
  \bibfield  {author} {\bibinfo {author} {\bibfnamefont {M.}~\bibnamefont
  {Fuchs}}\ and\ \bibinfo {author} {\bibfnamefont {M.~E.}\ \bibnamefont
  {Cates}},\ }\href {\doibase 10.1122/1.3119084} {\bibfield  {journal}
  {\bibinfo  {journal} {J. Rheol.}\ }\textbf {\bibinfo {volume} {53}},\
  \bibinfo {pages} {957} (\bibinfo {year} {2009})}\BibitemShut {NoStop}%
\bibitem [{\citenamefont {Succi}(2001)}]{Succi}%
  \BibitemOpen
  \bibfield  {author} {\bibinfo {author} {\bibfnamefont {S.}~\bibnamefont
  {Succi}},\ }\href@noop {} {\emph {\bibinfo {title} {The lattice Boltzmann
  equation for fluid dynamics and beyond}}}\ (\bibinfo  {publisher} {Oxford
  University Press},\ \bibinfo {address} {Oxford},\ \bibinfo {year}
  {2001})\BibitemShut {NoStop}%
\bibitem [{\citenamefont {D\"unweg}\ and\ \citenamefont
  {Ladd}(2009)}]{Duenweg.2009}%
  \BibitemOpen
  \bibfield  {author} {\bibinfo {author} {\bibfnamefont {B.}~\bibnamefont
  {D\"unweg}}\ and\ \bibinfo {author} {\bibfnamefont {A.~J.~C.}\ \bibnamefont
  {Ladd}},\ }\href@noop {} {\bibfield  {journal} {\bibinfo  {journal} {Adv.
  Polym. Sci.}\ }\textbf {\bibinfo {volume} {221}},\ \bibinfo {pages} {89}
  (\bibinfo {year} {2009})}\BibitemShut {NoStop}%
\bibitem [{\citenamefont {Papenkort}\ and\ \citenamefont
  {Voigtmann}(2014)}]{LBpaper}%
  \BibitemOpen
  \bibfield  {author} {\bibinfo {author} {\bibfnamefont {S.}~\bibnamefont
  {Papenkort}}\ and\ \bibinfo {author} {\bibfnamefont {{\relax
  Th}.}~\bibnamefont {Voigtmann}},\ }\href {\doibase
  http://dx.doi.org/10.1063/1.4872219} {\bibfield  {journal} {\bibinfo
  {journal} {J. Chem. Phys.}\ }\textbf {\bibinfo {volume} {140}},\ \bibinfo
  {eid} {164507} (\bibinfo {year} {2014}),\
  http://dx.doi.org/10.1063/1.4872219}\BibitemShut {NoStop}%
\bibitem [{\citenamefont {Giraud}\ \emph {et~al.}(1997)\citenamefont {Giraud},
  \citenamefont {d'Humi\`eres},\ and\ \citenamefont {Lallemand}}]{giraud97}%
  \BibitemOpen
  \bibfield  {author} {\bibinfo {author} {\bibfnamefont {L.}~\bibnamefont
  {Giraud}}, \bibinfo {author} {\bibfnamefont {D.}~\bibnamefont
  {d'Humi\`eres}}, \ and\ \bibinfo {author} {\bibfnamefont {P.}~\bibnamefont
  {Lallemand}},\ }\href@noop {} {\bibfield  {journal} {\bibinfo  {journal}
  {Int. J. Mod. Phys. C}\ }\textbf {\bibinfo {volume} {8}},\ \bibinfo {pages}
  {805} (\bibinfo {year} {1997})}\BibitemShut {NoStop}%
\bibitem [{\citenamefont {Giraud}\ \emph {et~al.}(1998)\citenamefont {Giraud},
  \citenamefont {D'Humi\`eres},\ and\ \citenamefont {Lallemand}}]{giraud98}%
  \BibitemOpen
  \bibfield  {author} {\bibinfo {author} {\bibfnamefont {L.}~\bibnamefont
  {Giraud}}, \bibinfo {author} {\bibfnamefont {D.}~\bibnamefont
  {D'Humi\`eres}}, \ and\ \bibinfo {author} {\bibfnamefont {P.}~\bibnamefont
  {Lallemand}},\ }\href@noop {} {\bibfield  {journal} {\bibinfo  {journal}
  {Europhys. Lett.}\ }\textbf {\bibinfo {volume} {42}},\ \bibinfo {pages} {625}
  (\bibinfo {year} {1998})}\BibitemShut {NoStop}%
\bibitem [{\citenamefont {Denniston}\ \emph {et~al.}(2001)\citenamefont
  {Denniston}, \citenamefont {Orlandini},\ and\ \citenamefont
  {Yeomans}}]{Denniston.2001}%
  \BibitemOpen
  \bibfield  {author} {\bibinfo {author} {\bibfnamefont {C.}~\bibnamefont
  {Denniston}}, \bibinfo {author} {\bibfnamefont {E.}~\bibnamefont
  {Orlandini}}, \ and\ \bibinfo {author} {\bibfnamefont {J.~M.}\ \bibnamefont
  {Yeomans}},\ }\href@noop {} {\bibfield  {journal} {\bibinfo  {journal} {Phys.
  Rev. E}\ }\textbf {\bibinfo {volume} {63}} (\bibinfo {year}
  {2001})}\BibitemShut {NoStop}%
\bibitem [{\citenamefont {Lallemand}\ \emph {et~al.}(2003)\citenamefont
  {Lallemand}, \citenamefont {d'Humi\`eres}, \citenamefont {Luo},\ and\
  \citenamefont {Rubinstein}}]{lallemand03}%
  \BibitemOpen
  \bibfield  {author} {\bibinfo {author} {\bibfnamefont {P.}~\bibnamefont
  {Lallemand}}, \bibinfo {author} {\bibfnamefont {D.}~\bibnamefont
  {d'Humi\`eres}}, \bibinfo {author} {\bibfnamefont {L.-S.}\ \bibnamefont
  {Luo}}, \ and\ \bibinfo {author} {\bibfnamefont {R.}~\bibnamefont
  {Rubinstein}},\ }\href {\doibase 10.1103/PhysRevE.67.021203} {\bibfield
  {journal} {\bibinfo  {journal} {Phys. Rev. E}\ }\textbf {\bibinfo {volume}
  {67}},\ \bibinfo {pages} {021203} (\bibinfo {year} {2003})}\BibitemShut
  {NoStop}%
\bibitem [{\citenamefont {Sulaiman}\ \emph {et~al.}(2006)\citenamefont
  {Sulaiman}, \citenamefont {Marenduzzo},\ and\ \citenamefont
  {Yeomans}}]{Sulaiman.2006}%
  \BibitemOpen
  \bibfield  {author} {\bibinfo {author} {\bibfnamefont {N.}~\bibnamefont
  {Sulaiman}}, \bibinfo {author} {\bibfnamefont {D.}~\bibnamefont
  {Marenduzzo}}, \ and\ \bibinfo {author} {\bibfnamefont {J.~M.}\ \bibnamefont
  {Yeomans}},\ }\href@noop {} {\bibfield  {journal} {\bibinfo  {journal} {Phys.
  Rev. E}\ }\textbf {\bibinfo {volume} {74}},\ \bibinfo {pages} {041708}
  (\bibinfo {year} {2006})}\BibitemShut {NoStop}%
\bibitem [{\citenamefont {Malaspinas}\ \emph {et~al.}(2010)\citenamefont
  {Malaspinas}, \citenamefont {Fi\'etier},\ and\ \citenamefont
  {Deville}}]{Malaspinas.2010}%
  \BibitemOpen
  \bibfield  {author} {\bibinfo {author} {\bibfnamefont {O.}~\bibnamefont
  {Malaspinas}}, \bibinfo {author} {\bibfnamefont {N.}~\bibnamefont
  {Fi\'etier}}, \ and\ \bibinfo {author} {\bibfnamefont {M.}~\bibnamefont
  {Deville}},\ }\href@noop {} {\bibfield  {journal} {\bibinfo  {journal} {J.
  Non-Newt. Fluid Mech.}\ }\textbf {\bibinfo {volume} {165}},\ \bibinfo {pages}
  {1637} (\bibinfo {year} {2010})}\BibitemShut {NoStop}%
\bibitem [{\citenamefont {Malaspinas}(2009)}]{Malaspinas}%
  \BibitemOpen
  \bibfield  {author} {\bibinfo {author} {\bibfnamefont {O.~P.}\ \bibnamefont
  {Malaspinas}},\ }\emph {\bibinfo {title} {Lattice Boltzmann Method for the
  Simulation of Viscoelastic Fluid Flows}},\ \href@noop {} {Ph.D. thesis},\
  \bibinfo  {school} {\'Ecole Polytechnique F\'ed\'erale de Lausanne} (\bibinfo
  {year} {2009})\BibitemShut {NoStop}%
\bibitem [{\citenamefont {Phillips}\ and\ \citenamefont
  {Roberts}(2011)}]{phillips11}%
  \BibitemOpen
  \bibfield  {author} {\bibinfo {author} {\bibfnamefont {T.~N.}\ \bibnamefont
  {Phillips}}\ and\ \bibinfo {author} {\bibfnamefont {G.~W.}\ \bibnamefont
  {Roberts}},\ }\href {\doibase 10.1093/imamat/hxr003} {\bibfield  {journal}
  {\bibinfo  {journal} {IMA J. Appl. Math.}\ }\textbf {\bibinfo {volume}
  {76}},\ \bibinfo {pages} {790} (\bibinfo {year} {2011})}\BibitemShut
  {NoStop}%
\bibitem [{\citenamefont {Benzi}\ \emph {et~al.}(2010)\citenamefont {Benzi},
  \citenamefont {Bernaschi}, \citenamefont {Sbragaglia},\ and\ \citenamefont
  {Succi}}]{Benzi.2010}%
  \BibitemOpen
  \bibfield  {author} {\bibinfo {author} {\bibfnamefont {R.}~\bibnamefont
  {Benzi}}, \bibinfo {author} {\bibfnamefont {M.}~\bibnamefont {Bernaschi}},
  \bibinfo {author} {\bibfnamefont {M.}~\bibnamefont {Sbragaglia}}, \ and\
  \bibinfo {author} {\bibfnamefont {S.}~\bibnamefont {Succi}},\ }\href@noop {}
  {\bibfield  {journal} {\bibinfo  {journal} {EPL}\ }\textbf {\bibinfo {volume}
  {91}},\ \bibinfo {pages} {14003} (\bibinfo {year} {2010})}\BibitemShut
  {NoStop}%
\bibitem [{\citenamefont {Sbragaglia}\ \emph {et~al.}(2012)\citenamefont
  {Sbragaglia}, \citenamefont {Benzi}, \citenamefont {Bernaschi},\ and\
  \citenamefont {Succi}}]{sbragaglia12}%
  \BibitemOpen
  \bibfield  {author} {\bibinfo {author} {\bibfnamefont {M.}~\bibnamefont
  {Sbragaglia}}, \bibinfo {author} {\bibfnamefont {R.}~\bibnamefont {Benzi}},
  \bibinfo {author} {\bibfnamefont {M.}~\bibnamefont {Bernaschi}}, \ and\
  \bibinfo {author} {\bibfnamefont {S.}~\bibnamefont {Succi}},\ }\href@noop {}
  {\bibfield  {journal} {\bibinfo  {journal} {Soft Matter}\ }\textbf {\bibinfo
  {volume} {8}},\ \bibinfo {pages} {10773} (\bibinfo {year}
  {2012})}\BibitemShut {NoStop}%
\bibitem [{\citenamefont {Benzi}\ \emph {et~al.}(2013)\citenamefont {Benzi},
  \citenamefont {Bernaschi}, \citenamefont {Sbragaglia},\ and\ \citenamefont
  {Succi}}]{Benzi.2013}%
  \BibitemOpen
  \bibfield  {author} {\bibinfo {author} {\bibfnamefont {R.}~\bibnamefont
  {Benzi}}, \bibinfo {author} {\bibfnamefont {M.}~\bibnamefont {Bernaschi}},
  \bibinfo {author} {\bibfnamefont {M.}~\bibnamefont {Sbragaglia}}, \ and\
  \bibinfo {author} {\bibfnamefont {S.}~\bibnamefont {Succi}},\ }\href@noop {}
  {\bibfield  {journal} {\bibinfo  {journal} {EPL}\ }\textbf {\bibinfo {volume}
  {104}},\ \bibinfo {pages} {48006} (\bibinfo {year} {2013})}\BibitemShut
  {NoStop}%
\bibitem [{\citenamefont {Marenduzzo}\ \emph {et~al.}(2007)\citenamefont
  {Marenduzzo}, \citenamefont {Orlandini}, \citenamefont {Cates},\ and\
  \citenamefont {Yeomans}}]{Marenduzzo.2007b}%
  \BibitemOpen
  \bibfield  {author} {\bibinfo {author} {\bibfnamefont {D.}~\bibnamefont
  {Marenduzzo}}, \bibinfo {author} {\bibfnamefont {E.}~\bibnamefont
  {Orlandini}}, \bibinfo {author} {\bibfnamefont {M.~E.}\ \bibnamefont
  {Cates}}, \ and\ \bibinfo {author} {\bibfnamefont {J.~M.}\ \bibnamefont
  {Yeomans}},\ }\href@noop {} {\bibfield  {journal} {\bibinfo  {journal} {Phys.
  Rev. E}\ }\textbf {\bibinfo {volume} {76}},\ \bibinfo {pages} {031921}
  (\bibinfo {year} {2007})}\BibitemShut {NoStop}%
\bibitem [{\citenamefont {Henrich}\ \emph {et~al.}(2010)\citenamefont
  {Henrich}, \citenamefont {Marenduzzo}, \citenamefont {Stratford},\ and\
  \citenamefont {Cates}}]{Henrich.2010}%
  \BibitemOpen
  \bibfield  {author} {\bibinfo {author} {\bibfnamefont {O.}~\bibnamefont
  {Henrich}}, \bibinfo {author} {\bibfnamefont {D.}~\bibnamefont {Marenduzzo}},
  \bibinfo {author} {\bibfnamefont {K.}~\bibnamefont {Stratford}}, \ and\
  \bibinfo {author} {\bibfnamefont {M.~E.}\ \bibnamefont {Cates}},\ }\href@noop
  {} {\bibfield  {journal} {\bibinfo  {journal} {Comp. Math. Appl.}\ }\textbf
  {\bibinfo {volume} {59}},\ \bibinfo {pages} {2360} (\bibinfo {year}
  {2010})}\BibitemShut {NoStop}%
\bibitem [{\citenamefont {Frantziskonis}(2011)}]{Frantziskonis.2011}%
  \BibitemOpen
  \bibfield  {author} {\bibinfo {author} {\bibfnamefont {G.~N.}\ \bibnamefont
  {Frantziskonis}},\ }\href@noop {} {\bibfield  {journal} {\bibinfo  {journal}
  {Phys. Rev. E}\ }\textbf {\bibinfo {volume} {83}},\ \bibinfo {pages} {066703}
  (\bibinfo {year} {2011})}\BibitemShut {NoStop}%
\bibitem [{\citenamefont {Su}\ \emph {et~al.}(2013)\citenamefont {Su},
  \citenamefont {Ouyang}, \citenamefont {Wang},\ and\ \citenamefont
  {Yang}}]{su13}%
  \BibitemOpen
  \bibfield  {author} {\bibinfo {author} {\bibfnamefont {J.}~\bibnamefont
  {Su}}, \bibinfo {author} {\bibfnamefont {J.}~\bibnamefont {Ouyang}}, \bibinfo
  {author} {\bibfnamefont {X.}~\bibnamefont {Wang}}, \ and\ \bibinfo {author}
  {\bibfnamefont {B.}~\bibnamefont {Yang}},\ }\href@noop {} {\bibfield
  {journal} {\bibinfo  {journal} {Phys. Rev. E}\ }\textbf {\bibinfo {volume}
  {88}},\ \bibinfo {pages} {053304} (\bibinfo {year} {2013})}\BibitemShut
  {NoStop}%
\bibitem [{\citenamefont {Keunings}(2003)}]{keunings03}%
  \BibitemOpen
  \bibfield  {author} {\bibinfo {author} {\bibfnamefont {R.}~\bibnamefont
  {Keunings}},\ }\href@noop {} {\bibfield  {journal} {\bibinfo  {journal}
  {Rheology Reviews}\ ,\ \bibinfo {pages} {167}} (\bibinfo {year}
  {2003})}\BibitemShut {NoStop}%
\bibitem [{\citenamefont {Tomé}\ \emph {et~al.}(2008)\citenamefont {Tomé},
  \citenamefont {de~Araujo}, \citenamefont {Alves},\ and\ \citenamefont
  {Pinho}}]{tome08}%
  \BibitemOpen
  \bibfield  {author} {\bibinfo {author} {\bibfnamefont {M.}~\bibnamefont
  {Tomé}}, \bibinfo {author} {\bibfnamefont {M.}~\bibnamefont {de~Araujo}},
  \bibinfo {author} {\bibfnamefont {M.}~\bibnamefont {Alves}}, \ and\ \bibinfo
  {author} {\bibfnamefont {F.}~\bibnamefont {Pinho}},\ }\href {\doibase
  http://dx.doi.org/10.1016/j.jcp.2007.12.023} {\bibfield  {journal} {\bibinfo
  {journal} {J. Comput. Phys.}\ }\textbf {\bibinfo {volume} {227}},\ \bibinfo
  {pages} {4207} (\bibinfo {year} {2008})}\BibitemShut {NoStop}%
\bibitem [{\citenamefont {Bergamasco}\ \emph {et~al.}(2013)\citenamefont
  {Bergamasco}, \citenamefont {Izquierdo},\ and\ \citenamefont
  {Ammar}}]{bergamasco13}%
  \BibitemOpen
  \bibfield  {author} {\bibinfo {author} {\bibfnamefont {L.}~\bibnamefont
  {Bergamasco}}, \bibinfo {author} {\bibfnamefont {S.}~\bibnamefont
  {Izquierdo}}, \ and\ \bibinfo {author} {\bibfnamefont {A.}~\bibnamefont
  {Ammar}},\ }\href@noop {} {\bibfield  {journal} {\bibinfo  {journal} {J.
  Non-Newt. Fl. Mech.}\ }\textbf {\bibinfo {volume} {201}},\ \bibinfo {pages}
  {29} (\bibinfo {year} {2013})}\BibitemShut {NoStop}%
\bibitem [{\citenamefont {Brader}\ \emph {et~al.}(2007)\citenamefont {Brader},
  \citenamefont {Voigtmann}, \citenamefont {Cates},\ and\ \citenamefont
  {Fuchs}}]{brader07}%
  \BibitemOpen
  \bibfield  {author} {\bibinfo {author} {\bibfnamefont {J.~M.}\ \bibnamefont
  {Brader}}, \bibinfo {author} {\bibfnamefont {{\relax Th}.}~\bibnamefont
  {Voigtmann}}, \bibinfo {author} {\bibfnamefont {M.~E.}\ \bibnamefont
  {Cates}}, \ and\ \bibinfo {author} {\bibfnamefont {M.}~\bibnamefont
  {Fuchs}},\ }\href {\doibase 10.1103/PhysRevLett.98.058301} {\bibfield
  {journal} {\bibinfo  {journal} {Phys. Rev. Lett.}\ }\textbf {\bibinfo
  {volume} {98}},\ \bibinfo {pages} {058301} (\bibinfo {year}
  {2007})}\BibitemShut {NoStop}%
\bibitem [{\citenamefont {Brader}\ \emph {et~al.}(2008)\citenamefont {Brader},
  \citenamefont {Cates},\ and\ \citenamefont {Fuchs}}]{brader08}%
  \BibitemOpen
  \bibfield  {author} {\bibinfo {author} {\bibfnamefont {J.~M.}\ \bibnamefont
  {Brader}}, \bibinfo {author} {\bibfnamefont {M.~E.}\ \bibnamefont {Cates}}, \
  and\ \bibinfo {author} {\bibfnamefont {M.}~\bibnamefont {Fuchs}},\ }\href
  {\doibase 10.1103/PhysRevLett.101.138301} {\bibfield  {journal} {\bibinfo
  {journal} {Phys. Rev. Lett.}\ }\textbf {\bibinfo {volume} {101}},\ \bibinfo
  {pages} {138301} (\bibinfo {year} {2008})}\BibitemShut {NoStop}%
\bibitem [{\citenamefont {Brader}\ \emph {et~al.}(2009)\citenamefont {Brader},
  \citenamefont {Voigtmann}, \citenamefont {Fuchs}, \citenamefont {Larson},\
  and\ \citenamefont {Cates}}]{brader09}%
  \BibitemOpen
  \bibfield  {author} {\bibinfo {author} {\bibfnamefont {J.~M.}\ \bibnamefont
  {Brader}}, \bibinfo {author} {\bibfnamefont {{\relax Th}.}~\bibnamefont
  {Voigtmann}}, \bibinfo {author} {\bibfnamefont {M.}~\bibnamefont {Fuchs}},
  \bibinfo {author} {\bibfnamefont {R.~G.}\ \bibnamefont {Larson}}, \ and\
  \bibinfo {author} {\bibfnamefont {M.~E.}\ \bibnamefont {Cates}},\ }\href
  {\doibase 10.1073/pnas.0905330106} {\bibfield  {journal} {\bibinfo  {journal}
  {Proc. Natl. Acad. Sci. U. S. A.}\ }\textbf {\bibinfo {volume} {106}},\
  \bibinfo {pages} {15186} (\bibinfo {year} {2009})}\BibitemShut {NoStop}%
\bibitem [{\citenamefont {Amann}\ \emph {et~al.}(2013)\citenamefont {Amann},
  \citenamefont {Siebenb\"urger}, \citenamefont {Kr\"uger}, \citenamefont
  {Weysser},\ and\ \citenamefont {Fuchs}}]{Amann.2013}%
  \BibitemOpen
  \bibfield  {author} {\bibinfo {author} {\bibfnamefont {C.~P.}\ \bibnamefont
  {Amann}}, \bibinfo {author} {\bibfnamefont {M.}~\bibnamefont
  {Siebenb\"urger}}, \bibinfo {author} {\bibfnamefont {M.}~\bibnamefont
  {Kr\"uger}}, \bibinfo {author} {\bibfnamefont {F.}~\bibnamefont {Weysser}}, \
  and\ \bibinfo {author} {\bibfnamefont {M.}~\bibnamefont {Fuchs}},\
  }\href@noop {} {\bibfield  {journal} {\bibinfo  {journal} {J. Rheol.}\
  }\textbf {\bibinfo {volume} {57}},\ \bibinfo {pages} {149} (\bibinfo {year}
  {2013})}\BibitemShut {NoStop}%
\bibitem [{\citenamefont {Siebenb\"urger}\ \emph {et~al.}(2012)\citenamefont
  {Siebenb\"urger}, \citenamefont {Ballauff},\ and\ \citenamefont
  {Voigtmann}}]{siebenbuerger12}%
  \BibitemOpen
  \bibfield  {author} {\bibinfo {author} {\bibfnamefont {M.}~\bibnamefont
  {Siebenb\"urger}}, \bibinfo {author} {\bibfnamefont {M.}~\bibnamefont
  {Ballauff}}, \ and\ \bibinfo {author} {\bibfnamefont {{\relax
  Th}.}~\bibnamefont {Voigtmann}},\ }\href {\doibase
  10.1103/PhysRevLett.108.255701} {\bibfield  {journal} {\bibinfo  {journal}
  {Phys. Rev. Lett.}\ }\textbf {\bibinfo {volume} {108}},\ \bibinfo {pages}
  {255701} (\bibinfo {year} {2012})}\BibitemShut {NoStop}%
\bibitem [{\citenamefont {Voigtmann}(2013)}]{ThVcreep}%
  \BibitemOpen
  \bibfield  {author} {\bibinfo {author} {\bibfnamefont {{\relax
  Th}.}~\bibnamefont {Voigtmann}},\ }\href {\doibase
  http://dx.doi.org/10.1063/1.4794555} {\bibfield  {journal} {\bibinfo
  {journal} {AIP Conf. Proc.}\ }\textbf {\bibinfo {volume} {1518}},\ \bibinfo
  {pages} {94} (\bibinfo {year} {2013})}\BibitemShut {NoStop}%
\bibitem [{\citenamefont {Larson}(1998)}]{Larson}%
  \BibitemOpen
  \bibfield  {author} {\bibinfo {author} {\bibfnamefont {R.~G.}\ \bibnamefont
  {Larson}},\ }\href@noop {} {\emph {\bibinfo {title} {The Structure and
  Rheology of Complex Fluids}}}\ (\bibinfo  {publisher} {Oxford University
  Press},\ \bibinfo {address} {Oxford},\ \bibinfo {year} {1998})\BibitemShut
  {NoStop}%
\bibitem [{\citenamefont {White}\ and\ \citenamefont
  {Metzner}(1963)}]{whiteMetzner63}%
  \BibitemOpen
  \bibfield  {author} {\bibinfo {author} {\bibfnamefont {J.~L.}\ \bibnamefont
  {White}}\ and\ \bibinfo {author} {\bibfnamefont {A.~B.}\ \bibnamefont
  {Metzner}},\ }\href@noop {} {\bibfield  {journal} {\bibinfo  {journal} {J.
  Appl. Polym. Sci.}\ }\textbf {\bibinfo {volume} {7}},\ \bibinfo {pages}
  {1867} (\bibinfo {year} {1963})}\BibitemShut {NoStop}%
\bibitem [{\citenamefont {Kim}\ and\ \citenamefont {Pitsch}(2007)}]{kim07}%
  \BibitemOpen
  \bibfield  {author} {\bibinfo {author} {\bibfnamefont {S.~H.}\ \bibnamefont
  {Kim}}\ and\ \bibinfo {author} {\bibfnamefont {H.}~\bibnamefont {Pitsch}},\
  }\href {\doibase 10.1063/1.2780194} {\bibfield  {journal} {\bibinfo
  {journal} {Phys. Fl.}\ }\textbf {\bibinfo {volume} {19}},\ \bibinfo {pages}
  {108101} (\bibinfo {year} {2007})}\BibitemShut {NoStop}%
\bibitem [{\citenamefont {{Palabos V1.1r0}}()}]{palabos}%
  \BibitemOpen
  \bibfield  {author} {\bibinfo {author} {\bibnamefont {{Palabos V1.1r0}}},\
  }\href@noop {} {}\bibinfo {howpublished}
  {\url{http://www.palabos.org}}\BibitemShut {NoStop}%
\bibitem [{\citenamefont {Nott}\ and\ \citenamefont {Brady}(1994)}]{Nott.1994}%
  \BibitemOpen
  \bibfield  {author} {\bibinfo {author} {\bibfnamefont {P.~R.}\ \bibnamefont
  {Nott}}\ and\ \bibinfo {author} {\bibfnamefont {J.~F.}\ \bibnamefont
  {Brady}},\ }\href@noop {} {\bibfield  {journal} {\bibinfo  {journal} {J.
  Fluid Mech.}\ }\textbf {\bibinfo {volume} {275}},\ \bibinfo {pages} {157}
  (\bibinfo {year} {1994})}\BibitemShut {NoStop}%
\bibitem [{\citenamefont {Batchelor}(1967)}]{Batchelor}%
  \BibitemOpen
  \bibfield  {author} {\bibinfo {author} {\bibfnamefont {G.~K.}\ \bibnamefont
  {Batchelor}},\ }\href@noop {} {\emph {\bibinfo {title} {An Introduction to
  Fluid Dynamics}}}\ (\bibinfo  {publisher} {Cambridge University Press},\
  \bibinfo {address} {Cambridge},\ \bibinfo {year} {1967})\BibitemShut
  {NoStop}%
\bibitem [{\citenamefont {Waters}\ and\ \citenamefont {King}(1970)}]{waters70}%
  \BibitemOpen
  \bibfield  {author} {\bibinfo {author} {\bibfnamefont {N.}~\bibnamefont
  {Waters}}\ and\ \bibinfo {author} {\bibfnamefont {M.}~\bibnamefont {King}},\
  }\href {\doibase 10.1007/BF01975401} {\bibfield  {journal} {\bibinfo
  {journal} {Rheol. Acta}\ }\textbf {\bibinfo {volume} {9}},\ \bibinfo {pages}
  {345} (\bibinfo {year} {1970})}\BibitemShut {NoStop}%
\bibitem [{\citenamefont {Huilgol}(2002)}]{Huilgol.2002}%
  \BibitemOpen
  \bibfield  {author} {\bibinfo {author} {\bibfnamefont {R.~R.}\ \bibnamefont
  {Huilgol}},\ }\href@noop {} {\bibfield  {journal} {\bibinfo  {journal} {Phys.
  Fl.}\ }\textbf {\bibinfo {volume} {14}},\ \bibinfo {pages} {1269} (\bibinfo
  {year} {2002})}\BibitemShut {NoStop}%
\bibitem [{\citenamefont {Huilgol}\ \emph {et~al.}(2002)\citenamefont
  {Huilgol}, \citenamefont {Mena},\ and\ \citenamefont {Piau}}]{Huilgol.2002b}%
  \BibitemOpen
  \bibfield  {author} {\bibinfo {author} {\bibfnamefont {R.~R.}\ \bibnamefont
  {Huilgol}}, \bibinfo {author} {\bibfnamefont {B.}~\bibnamefont {Mena}}, \
  and\ \bibinfo {author} {\bibfnamefont {J.~M.}\ \bibnamefont {Piau}},\
  }\href@noop {} {\bibfield  {journal} {\bibinfo  {journal} {J. Non-Newt. Fluid
  Mech.}\ }\textbf {\bibinfo {volume} {102}},\ \bibinfo {pages} {97} (\bibinfo
  {year} {2002})}\BibitemShut {NoStop}%
\bibitem [{\citenamefont {Chatzimina}\ \emph {et~al.}(2005)\citenamefont
  {Chatzimina}, \citenamefont {Georgiou}, \citenamefont {Argyropaidas},
  \citenamefont {Mitsoulis},\ and\ \citenamefont {Huilgol}}]{Chatzimina.2005}%
  \BibitemOpen
  \bibfield  {author} {\bibinfo {author} {\bibfnamefont {M.}~\bibnamefont
  {Chatzimina}}, \bibinfo {author} {\bibfnamefont {G.~C.}\ \bibnamefont
  {Georgiou}}, \bibinfo {author} {\bibfnamefont {I.}~\bibnamefont
  {Argyropaidas}}, \bibinfo {author} {\bibfnamefont {E.}~\bibnamefont
  {Mitsoulis}}, \ and\ \bibinfo {author} {\bibfnamefont {R.~R.}\ \bibnamefont
  {Huilgol}},\ }\href@noop {} {\bibfield  {journal} {\bibinfo  {journal} {J.
  Non-Newt. Fluid Mech.}\ }\textbf {\bibinfo {volume} {129}},\ \bibinfo {pages}
  {117} (\bibinfo {year} {2005})}\BibitemShut {NoStop}%
\end{thebibliography}%
\bibliographystyle{apsrev4-1}

\end{document}